\documentclass{aa}

\usepackage{epstopdf}
\usepackage{float}
\usepackage{graphicx}
\usepackage{txfonts}
\usepackage{xcolor}
\usepackage{CJK}
\usepackage{diagbox} 
\usepackage{graphicx}
\usepackage{booktabs}
\usepackage{array} 
\usepackage[pdftex,linkcolor=blue,citecolor=blue,backref=page,colorlinks=true]{hyperref}

\newcommand{\reffig}[1]{Figure~\ref{#1}}
\newcommand{\refeq}[1]{Equation~(\ref{#1})}

\newcommand{\Somb}{Sombrero-like galaxies}

\begin{document} 

   \title{Beyond Morphology: Challenges in Decomposing Massive Stellar Halos in Sombrero-like, Halo-Embedded Disk Galaxies}

   \authorrunning{He, Du, et al.}
   \titlerunning{Challenges in Decomposing Halo-Embedded Disk Galaxies}  
   \author{Wu-Tao He
          \inst{1}
          \and
          Min Du\inst{1}\fnmsep\thanks{Corresponding author: Min Du (Email: dumin@xmu.edu.cn)}
          \and 
          Zhao-Yu Li 
          \inst{2,3}
          \and
          Yuan Li
          \inst{4}
          }

   \institute{
    $^{1}$Department of Astronomy, Xiamen University, Xiamen, Fujian 361005, China\\
    $^{2}$Department of Astronomy, School of Physics and Astronomy, Shanghai Jiao Tong University, 800 Dongchuan Road, Shanghai 200240, China\\
    $^{3}$Key Laboratory for Particle Astrophysics and Cosmology (MOE) / Shanghai Key Laboratory for Particle Physics and Cosmology, Shanghai 200240, China\\
    $^{4}$Fujian Provincial Key Laboratory for Soft Functional Materials Research, Research Institute for Biomimetics and Soft Matter, Department of Physics, Xiamen University, Xiamen, Fujian 361005, China
}

  \abstract
   {}
   {Sombrero-like galaxies exhibit unique structural properties that challenge traditional photometric decomposition methods. We investigate the structural differences of Sombrero-like galaxies using both conventional photometric and kinematic decomposition approaches. This study aims to explore the extent to which photometric decomposition misidentifies key structural components, particularly the stellar halo.
}
   {We selected 270 Sombrero-like galaxies at redshift \( z = 0 \) from the TNG50 run of IllustrisTNG (TNG) simulations, filtering those with stellar mass \( M_\ast > 10^{10} M_\odot \) and stellar halo mass fraction satisfying \( 0.3 < f_{\rm halo} < 0.6 \). Synthetic images of these galaxies are generated using the GALAXEV population synthesis code, and photometric decomposition is performed on face-on and edge-on images using GALFIT. We then compare the decomposition results with kinematic decomposition based on the auto-GMM method, analyzing differences in the recovered structural parameters, including mass fractions and S\'ersic indices, and discussing their implications for identifying the bulge, disk, and stellar halo.  
}
   { Sombrero-like galaxies are characterized by disks embedded in massive stellar halos, namely halo-embedded disk galaxies. These galaxies likely represent 30–60\% of TNG50 galaxies, but identifying them is challenging due to structural degeneracies and the presence of disk features (e.g., bars, spirals, star formation) at low or moderate inclinations. Face-on photometric decomposition systematically overestimates disk fractions as stellar halos are almost absent, while edge-on analysis provides only approximate halo fractions. Radial profiles show discrepancies between photometric and kinematic decomposition, particularly in central regions. Additionally, No conclusive link exists between the S\'ersic index, $n$, and the presence of large stellar halos, challenging the use of $n$ as a merger history proxy. These findings underscore the need for improved decomposition methods to better understand the complex structures of Sombrero-like galaxies. The difficulty in identifying Sombrero-like galaxies, which have often undergone significant merger events, complicates our understanding of galaxy formation and evolution.}
    {}
   

   \maketitle

\section{Introduction}
The morphological classification and decomposition of galaxies have been widely used. As per the Hubble sequence \citep{1926ApJ....64..321H,1981rsac.book.....S}, galaxies can be categorized into types that are either dominated by rapidly rotating, flat disks or by pressure-supported, spherical bulges. The bulge is defined as a bright central concentration \citep{1936rene.book.....H}, a concept widely utilized in the visual classification of galaxies \citep[e.g.,][]{2013seg..book..155B}. The relative luminosity ratio of bulge to disk is usually determined by the morphological decomposition \citep[e.g.,][]{2002AJ....124..266P,2008A&A...478..353M,2015ApJ...799..226E,2019ApJS..244...34G}. This ratio directly decides the type of galaxy: early-type galaxies typically have a brighter bulge and quiescent star formation, while late-type galaxies have more disky morphology and ongoing star formation. The formation and evolution of the bulge and disk form the physical basis of the Hubble sequence. 

Growing evidence suggests that stellar halos are primarily the result of mergers, thus preserving fossil records of their hierarchical merging histories \citep[e.g.,][]{2016ApJ...821....5D,2018MNRAS.474.5300D,2019MNRAS.485.2589M, 2021ApJ...919..135D}. Bulges may have only a minimal connection to merger events. Numerical simulations have widely verified the dominant contribution of ex-situ stars to stellar halos, particularly in galaxies within the Milky Way mass range \citep{2009ApJ...702.1058Z,2010MNRAS.406..744C,2011MNRAS.416.2802F,2012MNRAS.420..255T,2019MNRAS.485.2589M,2022MNRAS.514.4898V,2023ApJ...943..158H}. Additionally, \citet{2024arXiv241203406X} suggested that the stellar mass in the outskirts of galaxies can effectively track the distribution of dark matter halos, perhaps due to the close relationship between the formation of dark matter halos and the establishment of stellar halos. Stellar halos possibly act as a fossil record of the formation and assembly histories of galaxies and their host dark matter halos. Despite their significance, it is a challenging task to measure stellar halos in external galaxies.

The low surface brightness of stellar halos \citep[e.g., \(\mu_V \ge 28 \, \text{mag/arcsec}^2\),][]{2005ApJ...635..931B} has made studying extragalactic halos a long-standing challenge. Deep imaging surveys (e.g., HST, Dragonfly) have revealed significant diversity in halo profiles and mass fractions \citep{2016ApJ...830...62M,2020ApJ...890...52C}. However, photometric methods often misestimate halos in transitional systems \citep{2012iac..talk..390S}, and stellar halos exhibit wide variations in density profiles, kinematics, and metallicity distributions \citep{2016MNRAS.457.1419M,2017MNRAS.466.1491H,2018MNRAS.475.3348H,2019ApJ...883..128G,2020ApJ...890...52C,2024ApJ...971..107O}. For example, NGC 4244's stellar halo has a modest mass of \(3 \times 10^6 M_\odot\) (just \(\sim 0.2\%\) of its total mass; \citep{2007ApJ...667L..49D,2015ApJS..219....5Q}, while M33’s halo constitutes a striking \(22 \pm 2\%\) of its mass based on RGB spectra \citep{2022ApJ...924..116G}—far exceeding photometric estimates \citep{2013MNRAS.428.1248C,2016MNRAS.461.4374M}. More massive galaxies show even greater diversity: M94 hosts two distinct RGB populations with a total accreted mass of \(2.8 \times 10^8 M_\odot\) \citep{2023ApJ...947...21G}; M104’s halo extends to 100 kpc \citep{2022ApJ...939...74K}; and NGC 5128 follows a steep (\(\propto r^{-3.1}\)) power-law density profile \citep{2022A&A...657A..41R}. The Dragonfly Telephoto Array found halo mass fractions ranging from 0.2\% to 6\% even among galaxies of similar mass and morphology, with no clear environmental dependence \citep{2016ApJ...830...62M,2020A&A...637A...8J}. M101, for instance, has a halo fraction of just \(0.20^{+0.10}_{-0.08}\%\). Extra-disc stellar components have also been detected in NGC 1560 and NGC 253 \citep{2014A&A...562A..73G,2018ApJ...861...81G}, and observations suggest a positive correlation between halo mass fraction and host galaxy mass \citep{2022ApJ...932...44G}. Intriguingly, some galaxies \citep[e.g., M101 in][]{2014ApJ...782L..24V,2016ApJ...830...62M}were initially thought to lack halos, but deeper HST imaging later identified a low-mass halo in M101 \citep{2020A&A...637A...8J}. This raises the critical question of whether these discrepancies are observational limitations, or whether stellar halos truly exhibit such irregular occurrence rates.

Both observations and simulations indicate that the Sombrero galaxy (M104) has massive stellar halos, which have long been regarded as a massive classical bulge. This leads to a different understanding. \citet{2012MNRAS.423..877G} proposed that when a stellar halo was included in their morphological fitting, the bulge's effective radius decreased to 0.46 kpc, its S\'ersic index dropped from \(n\sim4\) to 2, and the bulge-to-total (B/T) ratio decreased from 0.77 to 0.13. These changes are particularly remarkable when considering the scaling relations of different galaxy types. The bulge follows quite different scaling relations compared to the case without considering stellar halos. \citet{2020ApJ...895..139D} showed that a large number of Sombrero-like galaxies exist in the most advanced cosmological simulations, IllustrisTNG. These galaxies have extended spherical structures that are much larger than the typical scale of their disks. Sombrero-like galaxies are likely to be intermediate cases between elliptical and spiral galaxies that have been poorly studied. Such galaxies are likely to have experienced significant major mergers that destroyed their early disks and satellites but continued to form a new disk after the merger, as suggested by \citet{2021ApJ...919..135D}. 

In this study, Sombrero-like galaxies serve as an ideal sample for testing the effectiveness of morphological decomposition and highlighting the challenges they pose to galaxy classification within the Hubble sequence. Sombrero-like galaxies pose unique challenges for distinguishing classical bulges from stellar halos. Despite their relative abundance, they remain poorly studied. We systematically compare the kinematic and photometric decomposition methods for Sombrero-like galaxies in the TNG50 simulation. The structure of this paper is as follows. In Section~\ref{method}, we describe the methods for sample selection and data extraction. In Section~\ref{mcr}, we explore the diversity of morphological, color, and rotational properties of Sombrero-like galaxies. In Section~\ref{result}, we compare the results of morphological and kinematic decomposition. In Section~\ref{sec:discution}, we predict the observational probability of Sombrero-like galaxies at different redshifts and discuss the physical interpretation of the Sérsic index derived from photometric decomposition. Finally, in Section ~\ref{sm}, we summarize the main conclusions.

\section{Methodology}\label{method}

\begin{figure*}
       \centering
          \includegraphics[width=\textwidth]{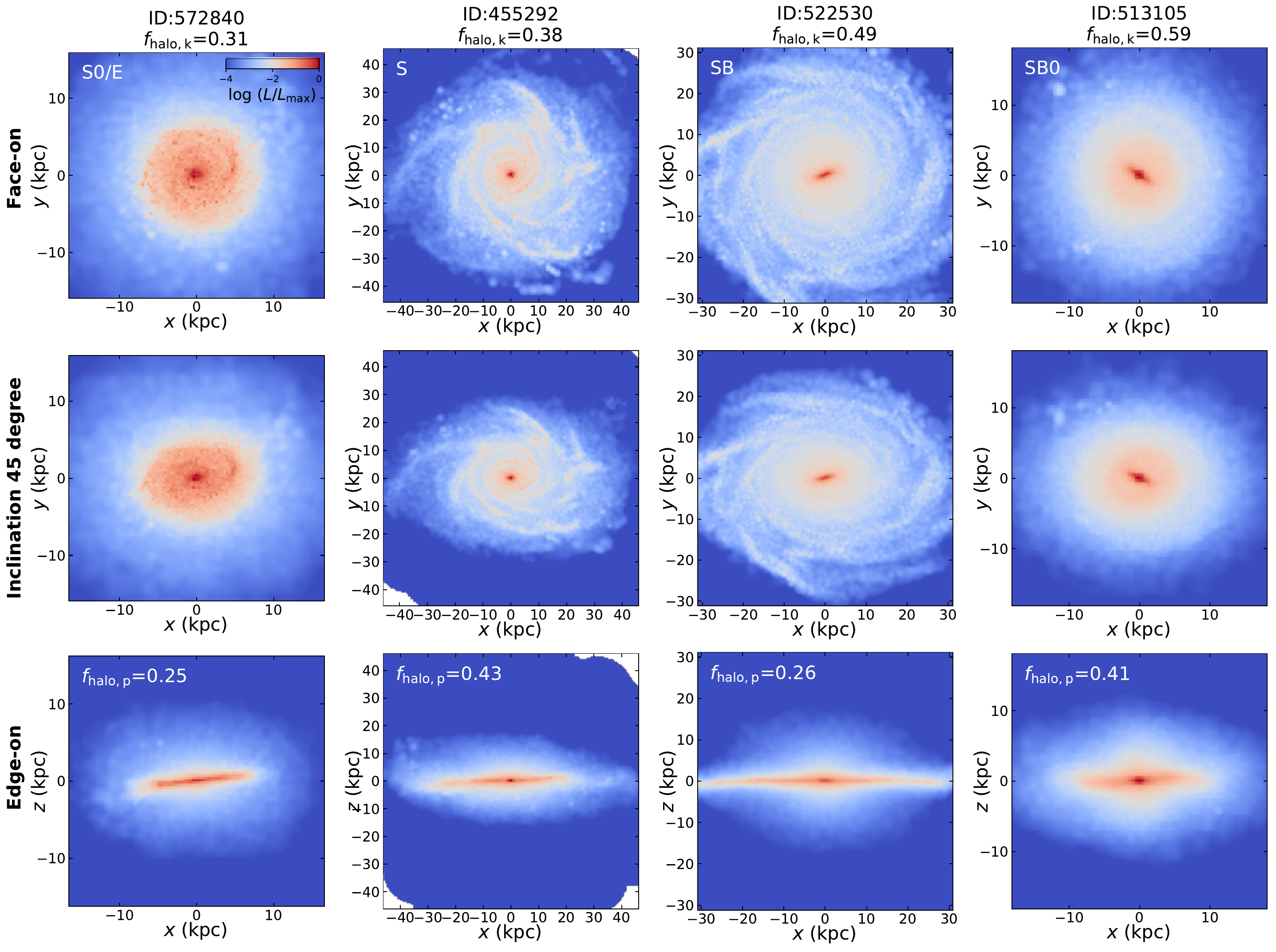}
       \caption{Mock images of Sombrero-like galaxies from the TNG50 simulations at z = 0, created using the {\tt GALAXEV} code. We present four galaxy examples (ID = 572840, 455292, 522530, 513105), where kinematically defined stellar halos account for 30-60\% of their total stellar mass ($\sim 10^{10.6} M_\odot$). Unlike observed Sombrero galaxies, the simulated Sombrero-like galaxies can be viewed at the same inclination angles. From top to bottom, the rows correspond to galaxy inclination angles of 0 degrees (face-on), 45 degrees, and 90 degrees (edge-on), with each image showing a field of view set to 7.5 times the half-mass radius from the galaxy center.
    }
                  \label{example}
        \end{figure*}

\subsection{The IllustrisTNG project}\label{TNG}

    Our analysis is based on one of the most advance cosmological simulations IllustrisTNG (hereafter referred to as TNG) \citep{2018MNRAS.475..624N,2019MNRAS.490.3234N,2018MNRAS.477.1206N,2018MNRAS.480.5113M,2018MNRAS.475..648P,2019MNRAS.490.3196P,2018MNRAS.475..676S}. This suite of simulations employs the moving-mesh code {\tt Arepo} \citep{2010MNRAS.401..791S} and $\Lambda$CDM cosmological parameters derived from \citet{2016ARA&A..54..529B}: $\Omega_m = 0.3089$, $\Omega_\Lambda = 0.6911$, $\Omega_b = 0.0486$, $h = 0.6774$, $\sigma_8 = 0.8159$, and $n_s = 0.9667$. The TNG model integrates crucial astrophysical processes including gas dynamics, fluctuations in the ultraviolet background, stellar formation, supernova (SN) feedback, stellar winds, interstellar medium enrichment, and the evolution of supermassive black holes. Galaxies are identified and characterized using the Friends-of-Friends \citep[FoF,][]{1985ApJ...292..371D} and {\tt SUBFIND} \citep{2001MNRAS.328..726S} algorithms. 
    
    The TNG simulations comprise three fiducial runs with approximate side lengths of 50, 100, and 300 Mpc, designated as TNG50, TNG100, and TNG300, respectively, each featuring distinct mass resolutions. Among these, TNG50  \citep{2019MNRAS.490.3196P,2019MNRAS.490.3234N} boasts the highest resolution initiated with $2 \times 2160^3$ elements. The mass of stellar particles reaches $8.5 \times 10^4 M_\odot$, and at $z = 0$, the gravitational softening length for stars is 0.3 kpc. It produces galaxies that possess realistic morphology and kinematics. \citet{2019MNRAS.490.3196P} showed that the typical thickness of star-forming galaxies in TNG50 is about a few hundred parsecs which matches well with observations. For galaxies with tiny kinematically inferred stellar halos, the resulting fiducial $j_\star-M_\star$-scale length (size)- metallicity relation is qualitatively consistent with observations \citep{Du2022, 2024A&A...686A.168D, Ma2024}. Both TNG50 and TNG100 simulations successfully reproduce bar structures \citep[e.g.,][]{Zhao2020, Lu2024}, which suggests that TNG simulations can accurately capture the internal dynamical processes of galaxies. Moreover, TNG50 can resolve many small-scale physical processes, such as cold gas clouds in the circumgalactic medium (CGM) stabilized by magnetic fields, with sizes of a few hundred parsecs \citep{2019MNRAS.490.3234N}, reasonable Milky Way-M31 pairs \citep{Pillepich2024}, small-scale gas structures around Milky Way-like galaxies \citep{2024MNRAS.528.3320R}, internal kinematics of dwarf satellites of MW/M31-like galaxies \citep{2023MNRAS.526.3589M}, and metallicity gradients \citep{2024A&A...684A..75C}.

 \begin{figure}
    \includegraphics[width=1
    \linewidth]{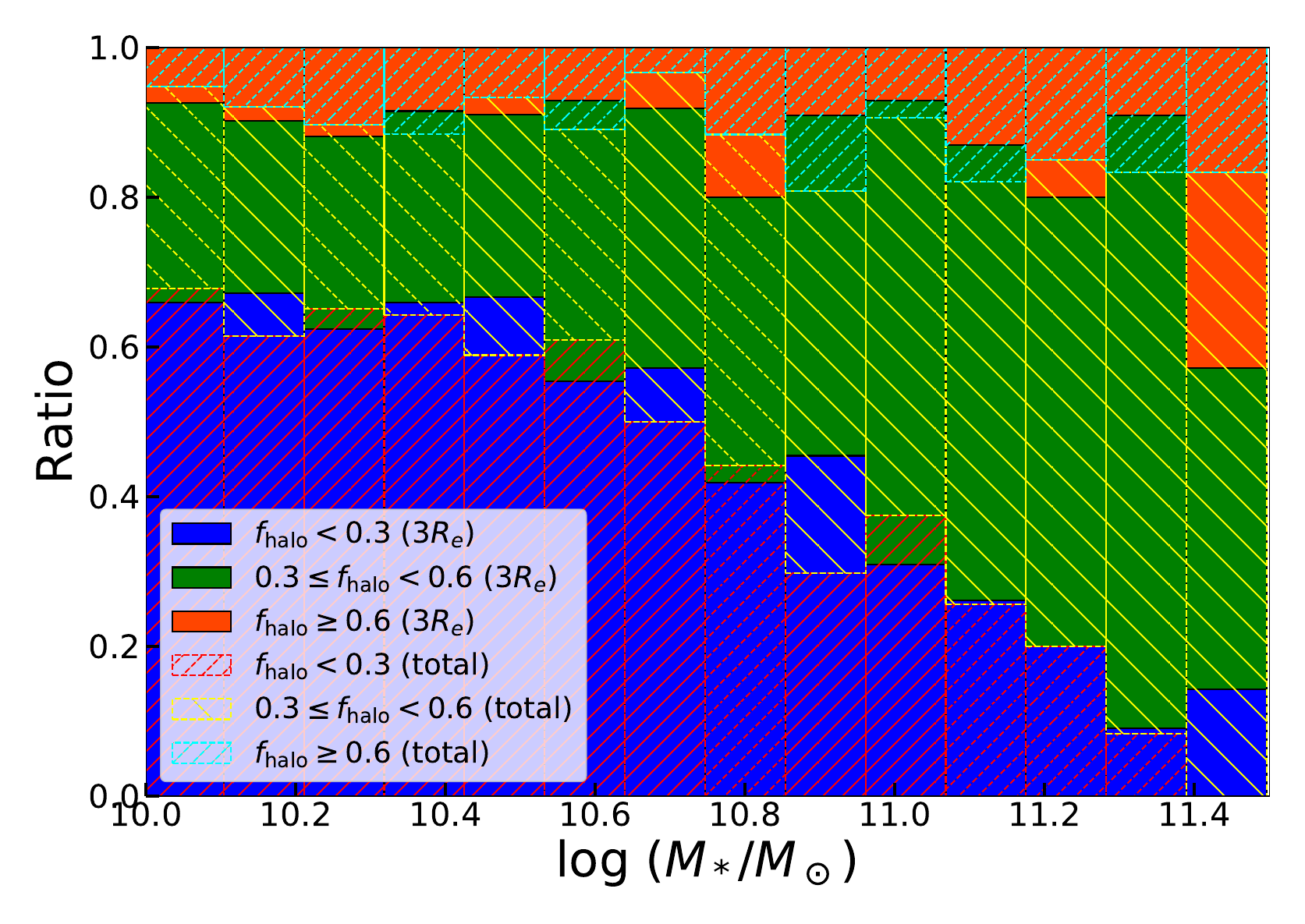}
    \caption{Galaxies from the TNG50 simulation are categorized into three ranges based on their stellar halo fraction $f_{\rm halo}$: $f_{\rm halo}<0.3$, $0.3\leq f_{\rm halo}<0.6$, and $f_{\rm halo}\geq0.6$. Galaxies with $0.3 \leq f_{\rm halo} < 0.6$ were selected as the sample of Sombrero-like galaxies. The proportion of galaxies within different mass intervals is shown. The ratio of stellar halo fraction is considered within a range of 3$R_e$ and the entire galaxy extent, represented by solid bar charts and dashed bar charts, respectively.
}
    \label{sample}
\end{figure}

\subsection{Sample selection of Sombrero-like, halo-embedded disk galaxies using kinematically defined stellar halos}\label{Sample}

    The Sombrero galaxy, known for its peculiarity, features a disk embedded in a massive and diffuse ``bulge.'' \citet{2012MNRAS.423..877G} suggested that considering an outer spheroidal component, akin to a stellar halo, could revise the bulge mass fraction from 77\% to about 10\% in the Sombrero galaxy. The existence of a large stellar halo can, therefore, pose significant challenges when measuring galaxy structures using morphological methods. Such Sombrero-like galaxies thus can also be named as halo-embedded disk galaxies, representing a transitional form between pure disk and elliptical galaxy morphologies.
    
    \citet{2019ApJ...884..129D} recently developed an automated framework, they called {\tt auto-GMM}, to efficiently decompose the structures of simulated galaxies based on their kinematic phase-space properties. This framework identifies components such as disks, bulges, and stellar halos \citep{2020ApJ...895..139D} by clustering stars according to their positions in a three-dimensional phase space defined by circularity, binding energy, and non-azimuthal angular momentum \citep{2003ApJ...597...21A,2012MNRAS.421.2510D} in a physically significant and unbiased manner. Stars grouped into kinematically defined disks are primarily characterized by strong rotational motion, with a mass-weighted mean circularity $\langle j_z/j_{\rm c}\rangle > 0.5$, where $j_z$ and $j_{\rm c}$ represent azimuthal and circular angular momentum, respectively. In contrast, kinematically derived stellar halos share the trait of weak rotation ($\langle j_z/j_{\rm c}\rangle < 0.5$) with bulges but contain less tightly bound stars than bulges. Both bulges and stellar halos exhibit spheroidal morphologies. Bulges mainly comprise stars that are strongly gravitationally bound within the central regions of galaxies, whereas stellar halos represent extended and loosely bound spheroidal structures, significantly shaped by ``dry'' mergers that disrupt the disks of their progenitors and satellites merging in \citep{2021ApJ...919..135D}. In this study, $f_{\rm halo}, f_{\rm bulge}$, and $f_{\rm disk}$ are defined as the mass fractions of the kinematically derived stellar halo, bulge, and disk in each galaxy, respectively. The data are publicly accessible\footnote{The data of kinematic structures in TNG galaxies are publicly accessible at https://www.tng-project.org/data/docs/specifications/\#sec5m}. 

    A group of Sombrero-like galaxies are identified in TNG simulations. The kinematic decomposition of TNG galaxies using {\tt auto-GMM} reveals that many galaxies possess a significant stellar halo component, often mistakenly classified as bulges or disks in morphology \citep{2020ApJ...895..139D}. In this study, we curated a collection of Sombrero-like galaxies; for visual reference, consult Figure \ref{example} herein and Figure 5 of \citet{2021ApJ...919..135D} for instance, where a disk component is embedded in a massive stellar halo. We identify 270 Sombrero-like TNG50 galaxies with stellar masses log \((M_\star/M_\odot) > 10\) measured within three times of their half-mass radius $R_e$. Their stellar halos $f_{\rm halo}$ derived by {\tt auto-GMM} vary from 0.3 to 0.6, as shown in Figure \ref{example} from left to right. Figure \ref{sample} shows that the Sombrero-like galaxies we select populate widely in a wide range of stellar masses. Figure \ref{sample} depicts the fractions of typical elliptical galaxies characterized by $f_{\rm halo} > 0.6$ (orange), disk galaxies $f_{\rm halo} < 0.3$ (blue), and Sombrero-like galaxies (green), all measured within a radius of $r<3 {r_e}$, where \( r_e \) is the effective radius derived from 3D polar coordinates, at redshift \( z = 0 \). For comparison, the results measured without any radius constraints are also shown. It is clear that Sombrero-like galaxies typically constitute about 20-30\% of galaxies with log $(M_\star/M_\odot)<10.7$, but escalating to approximately 40-60\% toward more massive cases. 

\subsection{Generation of mock images using {\tt SKIRT} and {\tt GALAXEV}}\label{mockimage}

    Dust absorbs and disperses the light from galaxies, especially in the optical band. The luminosity of galaxies thus is likely to be underestimated, especially in the central regions of galaxies. Consequently, it may result in great uncertainty in the morphological decomposition of galaxy structures. In this study, we utilize mock images generated by the three-dimensional radiative transfer code SKIRT \citep{2015A&C.....9...20C,2020A&C....3100381C} and those generated by {\tt GALAXEV} where detailed spatial variations of dust are not taken into account. The cases without considering the effect of dust are used for comparison. The impact of a large uncertainty of dust models \citep[e.g.,][]{Rodriguez-Gomez2019} then is examined. By comparing the results from SKIRT and GALAXEV, we ensure the robustness and reliability of our findings. Since stellar halos are extended spheroidal stellar components that are mostly measured in edge-on views where there is no dust. The choice of a different dust model has a minor impact on our main conclusions. However, it does affect the measurement of bulges to some extent.

    We adopt the TNG50-SKIRT Atlas\footnote{The data of TNG50-SKIRT Atlas is publicly available at https://www.tng-project.org/data/docs/specifications/\#sec5\_7} \citep[{\tt TSA},][]{2024A&A...683A.182B}, which is an extensive multiwavelength synthetic image repository based on simulated galaxies from TNG50. For each simulated galaxy, {\tt TSA} performs SKIRT. Originally designed as a dust radiative transfer tool for simulating the effects of dust in galaxies, SKIRT has evolved into a more versatile Monte Carlo radiative transfer tool. All images in TSA are generated with SKIRT, which considers various stellar populations and interstellar dust absorption and scattering in realistic 3D environments. The adopted SKIRT simulation methodology is based on the prescriptions of \citet{2016MNRAS.462.1057C,2018ApJS..234...20C,2022MNRAS.512.2728C}, \citet{2021MNRAS.506.5703K}, and \citet{2022MNRAS.516.3728T}. For each galaxy, the primary radiation sources are the stellar particles belonging to the galaxy, extracted from the TNG50 database. For each stellar particle older than 10 Myr, a simple stellar population (SSP) spectral energy distribution (SED) is established from the \citet{2003MNRAS.344.1000B} SSP family, employing a \citet{2003PASP..115..763C} initial mass function. Stellar particles younger than 10 Myr are assigned SEDs based on the HII region templates from the MAPPINGS III library \citep{2008ApJS..176..438G}. The dust grain model uses the diffuse ISM THEMIS model \citep{2017A&A...602A..46J}, which consists of a distribution of carbonaceous and silicate grains, incorporating, where possible, optical properties based on laboratory data. The images are generated from five fixed viewpoints relative to the simulation box, independent of each galaxy's intrinsic orientation. The image atlas includes images from five randomized observer positions for each galaxy. All images and parameter maps are 1600 × 1600 pixels, with a pixel scale of 100 pc, corresponding to a field of view of 160 kpc. We use mock LSST $r$-band images from the TNG50-SKIRT dataset. By conducting elliptical fitting on images generated from five random viewing angles, we regard the cases with the smallest overall ellipticity as the face-on view and the one with the largest overall ellipticity as the edge-on view.

    Although {\tt TSA} applies a much more realistic radiative transfer model, it does not have accurate face-on and edge-on measurements. We thus adopt the stellar population synthesis code, designated as {\tt GALAXEV} \citep{2003MNRAS.344.1000B}, to create mock images of Sombrero-like galaxies from TNG simulations. Similar results are obtained by these two models. The stellar spectra models used in this process were derived from the Padova 1994 evolutionary tracks. This study focuses on the degeneracy effect between stellar halos and other galactic structures. We adopt a simply spatially unresolved dust distribution model, as per \citep{2000ApJ...539..718C}. This dust model primarily relies on an idealized description of dust and radiation within galaxies. It assumes that dust is uniformly mixed throughout the galaxy, without considering detailed spatial variations. This approach models the absorption and scattering of stellar radiation by dust to infer its effect on the integrated spectral properties of galaxies. We incorporate this model to assess the potential impact of dust extinction on the morphological appearance of galaxies. We opt not to utilize the radiative transfer model in order to simplify our analysis of the degeneracy among various structures. This approach helps us to streamline our study by avoiding unnecessary complexity that the model might introduce, allowing us to focus more directly on the interplay between different structural elements. Dust extinction should be least significant on the measurements of stellar halos in galaxies observed in nearly edge-on views, an orientation that is particularly advantageous for studying stellar halos. The result in the case that no dust extinction is considered is also examined for comparison. The telescope we consider for comparison is the Hyper Suprime-Cam (HSC) for the Subaru Telescope. Therefore, we adjust the instrument parameters to match those of HSC (0.168 arcsec/pix, 0.318 kpc/pix). To ensure that the galaxy sizes in the images are roughly comparable and fill the frame, we standardize the image rendering by setting the 7.5 times the half-mass radius from the galaxy center. We simulated observations using the $g$, $r$, $i$, $z$, and $y$ filters, which are commonly used in actual observations. We then project the galaxies to different orientations. Finally, we synthesized mock images of the TNG50 galaxies selected according to our criteria, matching the HSC specifications.

\begin{figure}
	\includegraphics[width=0.9\linewidth]{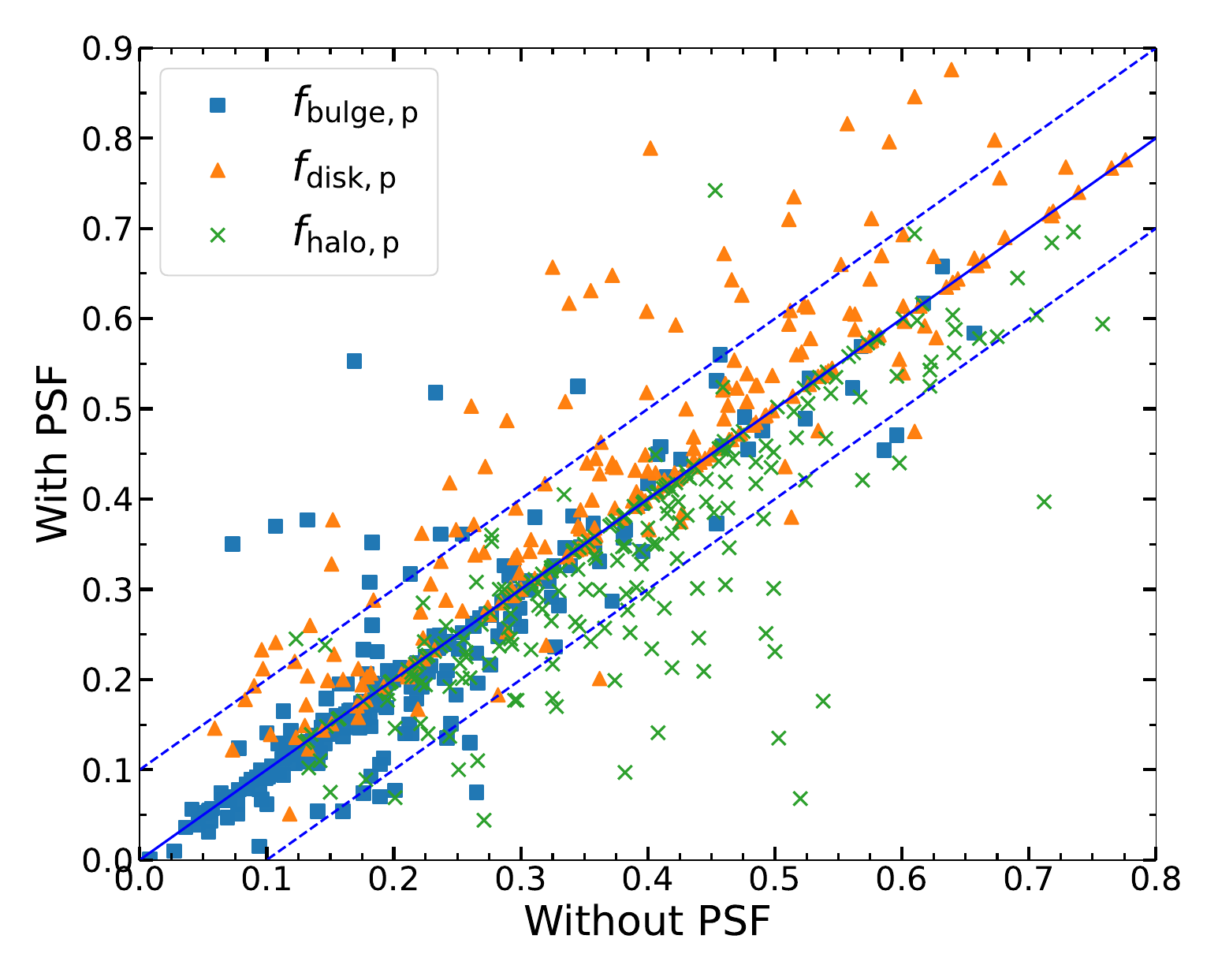}
    \caption{Relation between photometric component fractions derived with and without PSF convolution in edge-on synthetic images. Each point represents a Sombrero-like galaxy. Circles, triangles, and squares correspond to the ratio of photometrically measured bulges ($f_\mathrm{bulge,p}$), disks ($f_\mathrm{disk,p}$), and stellar halos ($f_\mathrm{halo,p}$), respectively. The blue solid line denotes the ideal 1:1 correspondence, while the dashed lines indicate a $\pm 0.1$ deviation range.
}
    \label{psf}
\end{figure}

    We created mock images of all the selected Sombrero-like galaxies in face-on, 45-degree inclined, and edge-on orientations to consider the influence of galaxy inclination on photometric decomposition. These simulated images were saved as FITS files. Figure \ref{example} presents mock images of four examples measured in their $r$-band from {\tt GALAXEV}. It should be noted that neither the Point Spread Function (PSF) nor background contributions are included in our simplified analysis. This simplification is driven by the fact that structural degeneracies (e.g., overlapping features like halos or disks) present a more substantial analytical challenge compared to instrumental effects. As shown in Figure \ref{psf}, tests with a 2D Gaussian PSF (FWHM = 0.7" , pixel scale = 0.168"/pix) generally show minimal impact on stellar halo fractions measured in morphology, with discrepancies below 0.1. Our findings can therefore be regarded to be the ideal case.And we verified that the results obtained using TSA images show no significant differences from those derived using GALAXEV.

\subsection{Morphological decomposition of Sombrero-like, halo-embedded disk galaxies}\label{sec:Photometric decomposition}

    We performed photometric decomposition on the mock $r$-band images of Sombrero-like galaxies from both {\tt TSA} and {\tt GALAXEV}, as introduced in Section \ref{mockimage}, using {\tt GALFIT} \citep{2002AJ....124..266P}. It's well known that securing high-quality, multi-component fits for extensive datasets is challenging. To mitigate the incidence of critical failures during the {\tt GALFIT} fitting process, we implemented a preliminary step where each pixel's value in the image was transformed from flux units to count units prior to fitting. This conversion was crucial in preventing errors arising from extremely small flux values. Moreover, the use of count units facilitates the internal generation of sigma images in {\tt GALFIT}. By working in count units, we enhance fitting precision, streamline data processing, and bolster the reliability of our analytical outcomes.

    \begin{figure*}
       \centering
          \includegraphics[width=0.8\textwidth]{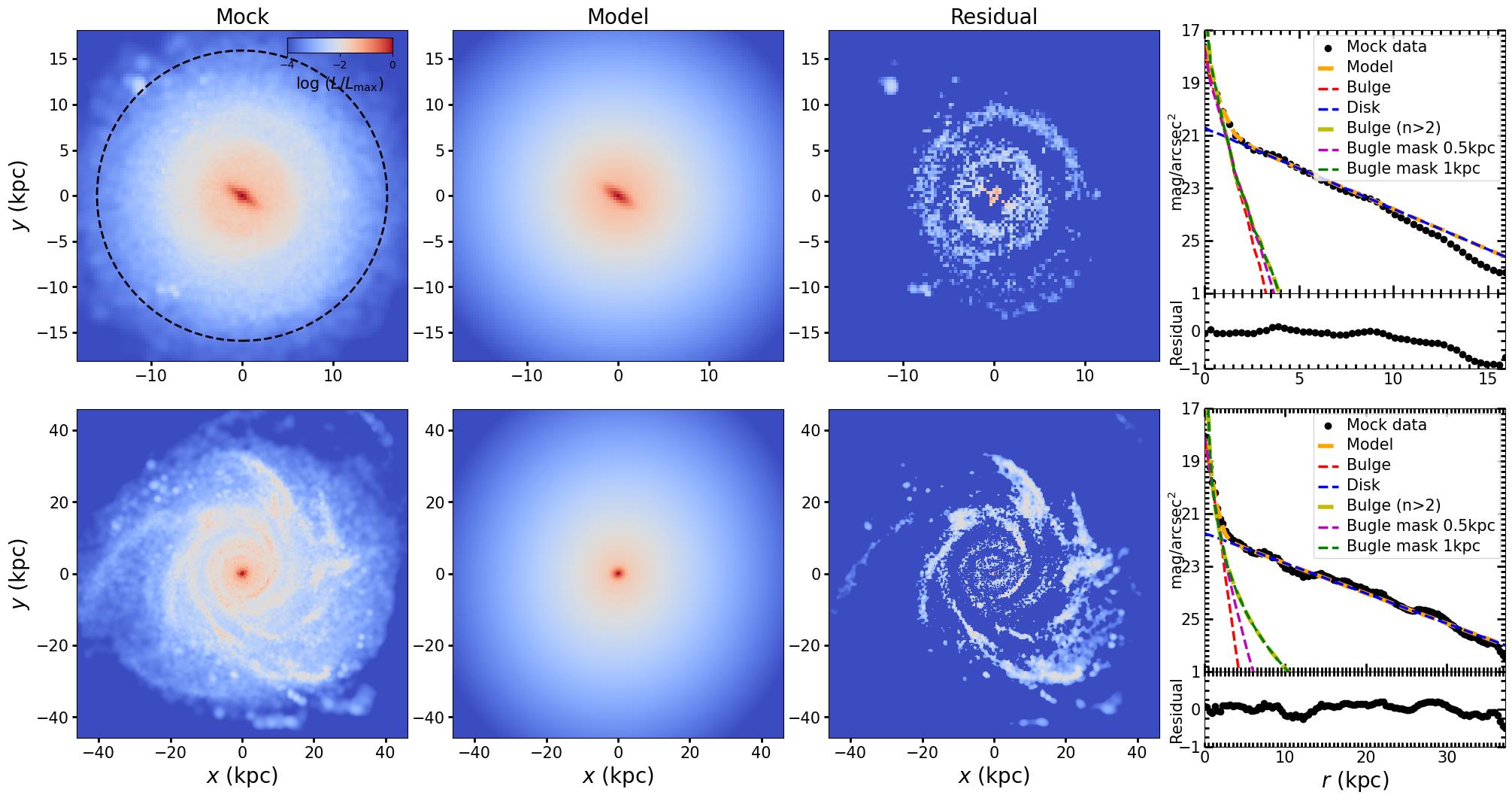}
       \caption{Example of photometric decomposition for a face-on galaxy (ID=513105, 455292). The first column shows the face-on mock image generated through the {\tt GALAXEV} pipeline. The dashed black line represents a circle drawn at the average radius where the pixel value equals 1, distinguishing the background from the stellar halo. The second column presents the two-component {\tt GALFIT} model (bulge + disk). The third column displays the residuals, where the pixel values are normalized by the maximum value of each galaxy. The fourth column shows the surface brightness profile fitting results.
    }
                  \label{faceon}
        \end{figure*}  

 \begin{figure*}
       \centering
          \includegraphics[width=0.8\textwidth]{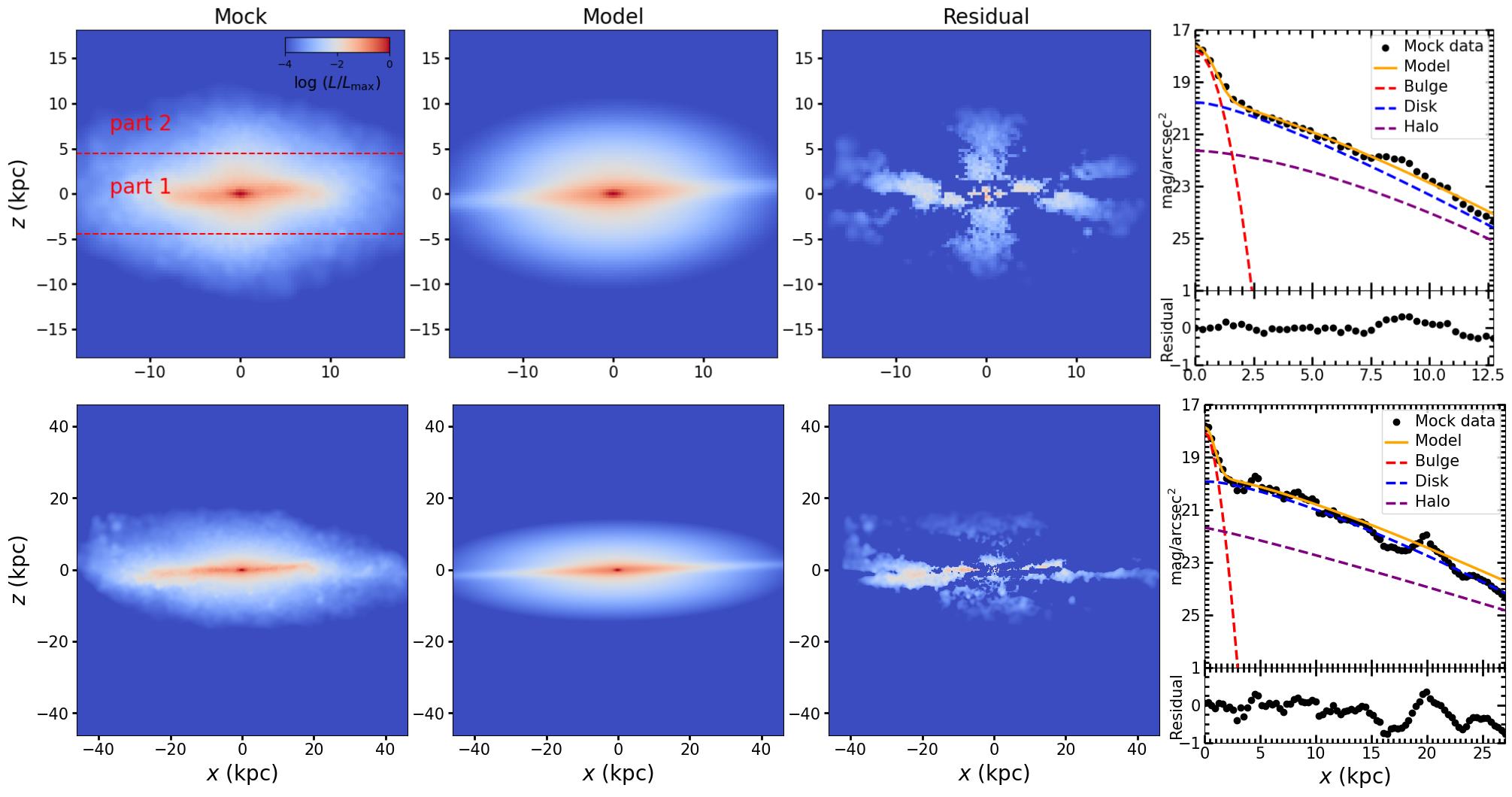}
       \caption{Examples of photometric decomposition for galaxies in the edge-on view (ID=513105, 455292). The first column shows the edge-on mock image generated using the {\tt GALAXEV} pipeline. The dashed red line, located 1Re from the galaxy center, divides the edge-on galaxy into part 1 and part 2. The second column presents the three-component (bulge+disk+stellar halo) model fit by {\tt GALFIT}. The third column displays the residual of this fitting result. The fourth column shows the one-dimensional surface brightness of different components along the major axis that is oriented to be aligned with the $x$ axis.
    }
                  \label{edgeon}
        \end{figure*}

\subsubsection{Photometric decomposition of disks, bulges, and stellar halos in face-on and edge-on views}

We performed two-component and three-component galaxy fitting for both face-on and edge-on galaxies. We assume that the surface brightness of the bulge and halo follows the S\'ersic profile \citep{1963BAAA....6...41S}.

We assume that the surface brightness of the disk in the face-on direction follows the exponential function
$I(R) = I_0 \exp \left( -\frac{R}{h_R} \right)$
where $I_0$ and $h_R$ are the central surface brightness and scale length, respectively. There is a relation between $R_{\text{e,disk}}$ and $h_R$ given by $R_{\text{e,disk}} = 1.678h_R$.
The edge-on surface density of edge-on disks is described by
\begin{equation}
I(R, z) = I_0 \left( \frac{R}{h_R} \right) K_1 \left( \frac{R}{h_R} \right) \text{sech}^2 \left( \frac{z}{h_z} \right),
    \label{eq-Disk Profile}
\end{equation}where $h_z$ and $K_1$ are scale height and the Bessel function, respectively.

{\tt GALFIT} uses the Levenberg-Marquardt algorithm to find the optimal solution by minimizing $\chi^2$. We ran {\tt GALFIT} multiple times with a range of initial conditions for each galaxy, as follows, to reduce the likelihood of converging on a solution that only represents a local $\chi^2$ minimum.
\begin{itemize}
    \item S\'ersic index: $n = 1, 2, 4$
    \item Bulge half-light radius: $R_{\text{e,bulge}} = 0.2, 0.5 \times R_e$
    \item Halo half-light radius: $R_{\text{e,halo}} = 1, 2, 3 \times R_e$   
    \item $h_R = 1, 1.5, 2 \times R_e$   
    \item $h_z = 0.2, 0.5 \times R_e$
\end{itemize}
None of these parameters are fixed while the fitting is running.

In face-on or 45$^\circ$ views, a two-component S\'ersic bulge+exponential disk model can fit TNG50 galaxies well, so there is no need to consider the stellar halo component.
Nevertheless, in edge-on views of the same sample of galaxies, a three-component model, namely bulge, disk, and stellar halo, performs much better (as shown in Figure \ref{edgeon}). This is partly because the region overlapping between the disk and the halo along the line of sight becomes small, making the presence of a significant stellar halo component more apparent. As suggested in \citet{2012MNRAS.423..877G}, a third S\'ersic component that corresponds to a stellar halo is required to fit edge-on Sombrero-like galaxies accurately.

To isolate the stellar halos of edge-on galaxies and prevent local minima during multi-component fitting, we mask the inner regions of these galaxies. As illustrated in the upper left panel of Figure \ref{edgeon}, the masked inner regions are outlined by dashed red lines (referred to as Part 1). The height of the masked region varies, covering 0.5$R_e$, 1$R_e$, and 1.5$R_e$, and the appropriate mask width is selected based on the fitting results. Part 1 consists of the bulge, disk, and stellar halo components, while Part 2 is primarily dominated by the stellar halo. The fitting process proceeds as follows: First, we performed a S\'ersic + edge-on disk fit (see \refeq{eq-Disk Profile}) for Part 1. Next, we fit Part 2, which includes only the S\'ersic stellar halo. Finally, we use the results from these fits to conduct a three-component fit for the entire edge-on galaxy. The fitting results, as demonstrated in Figure \ref{edgeon}, show excellent agreement between the mock image and the radial profile of the fit model. 
Hereafter, $f_\text{bulge,p}$, $f_\text{disk,p}$, $f_\text{residual,p}$, and $f_\text{halo,p}$ represent the luminosity ratios of the bulge, disk, stellar halo, and residual components, respectively, in the morphological decomposition. These ratios are used to compare with kinematically derived structures.

\begin{figure}
	\includegraphics[width=0.8\linewidth]{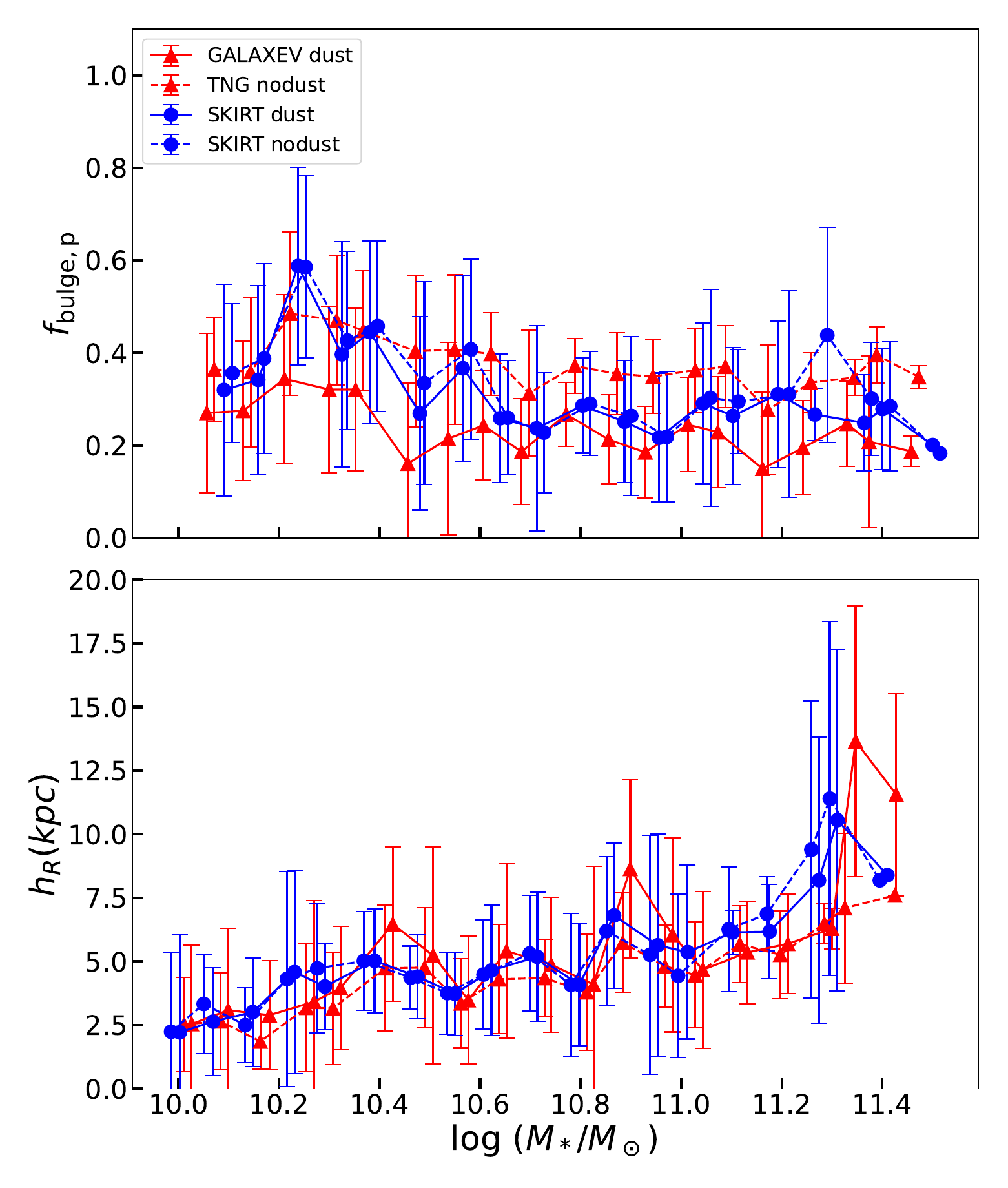}
    \caption{{\tt GALFIT} fitting results using mock images generated by {\tt GALAXEV} and {\tt SKIRT}, in both dust-free (dashed lines) and dusty (solid lines) conditions. The upper panel shows the variation in bulge fraction $f_{\text{bulge}, p}$ with stellar mass. The lower panel illustrates how the scale height of the disk $h_R$. Dashed and solid lines represent the median values, whose error bars correspond to the $1\sigma$ range. 
}
    \label{ny_dust}
\end{figure}

We confirm that mock images generated by {\tt GALAXEV} and {\tt SKIRT} yield consistent results. The upper panel of Figure \ref{ny_dust} displays how the disk scale height \( h_R \) changes with galaxy stellar mass \( \log(M_*/M_\odot) \). The curves represent the median \( h_R \) of disk components across different mass ranges, with error bars indicating the one-sigma range. The lower panel shows the bulge fraction \( f_{\rm bulge, p} \) as a function of galaxy stellar mass, with curves representing the median \( f_{\rm bulge, p} \) in different mass ranges, and error bars indicating the one-sigma range. While galaxy sizes remain similar in both dusty and dust-free cases, the effect of dust extinction causes the decomposed images of dusty galaxies to appear more diffuse compared to those without dust. The bulge fraction derived from the photometric decomposition of dust-free galaxies is approximately 0.1 higher than that of dusty galaxies. This difference occurs because dust alters the luminosity distribution of galaxies, smoothing the brightness profile and leading to an underestimation of the bulge contribution while overestimating the disk contribution.

\subsubsection{Quantifying fit quality: Residuals and $\chi^2$ statistics}\label{evaluate}

Figures \ref{faceon} and \ref{edgeon} display the relative galaxy surface luminosity \(L\) that has been normalized by the maximum value \(L_{\rm max}\). From left to right, the mock images, fitting results, and residual images are presented. In order to avoid excessive computation from influencing the statistical results, we set a threshold of pixel values greater than 1 (equivalent to an AB magnitude of 26 mag) as the region of interest. This threshold effectively distinguishes the stellar halo from the background. Both the radial profile and relative error calculations are performed within this region. The right-most panels of Figures \ref{faceon} and \ref{edgeon} display the one-dimensional surface density distribution of each component and the residual. We then assess the goodness of fit by radial profile images, residual images, and $\chi^2$. 260 galaxies are successfully fit in their face-on orientation, and 242 are well fit in the edge-on orientation. In visual inspection, seven galaxies have irregular morphology and were thus excluded from our sample. Additionally, four bulgeless galaxies can be well-fit using two components, i.e., a S\'ersic stellar halo and an exponential disk.

   We calculated the $\chi^2$ values and the distribution of relative error values of the fitting image along both the major and minor axes. Most galaxies exhibit small $\chi^2$ values and relative errors less than 0.3. Similarly, edge-on galaxies generally have slightly lower $\chi^2$ values compared to face-on galaxies. In this study, we mainly draw conclusions based on cases where the relative errors along both the major and minor axes are less than 0.3, in order to exclude the potential influence of poorly fit galaxies. Cases with larger relative errors are also presented to ensure the completeness of our sample.

\subsubsection{Excluding the potential influence of thick disks on our results}

    There is no distinct feature to tell thick disks and stellar halos apart in morphology, as seen in the second column of Figure \ref{example}. In our selected Sombrero-like galaxies, we classify those galaxies that have a thick disk component, where the ratio of the minor axis ($b$) to the major axis ($a$) at the effective radius, quantified through the GALFIT fitting of the stellar halo, is less than 0.3. Most of stellar halos have $b/a$ varying between $0.3$ and $0.7$. 47 galaxies are classified as having a thick disk structure. This matches our visual inspection. We have verified that the existence of the thick disk has no significant impact on our results.

\begin{figure}
	\includegraphics[width=0.9\linewidth]{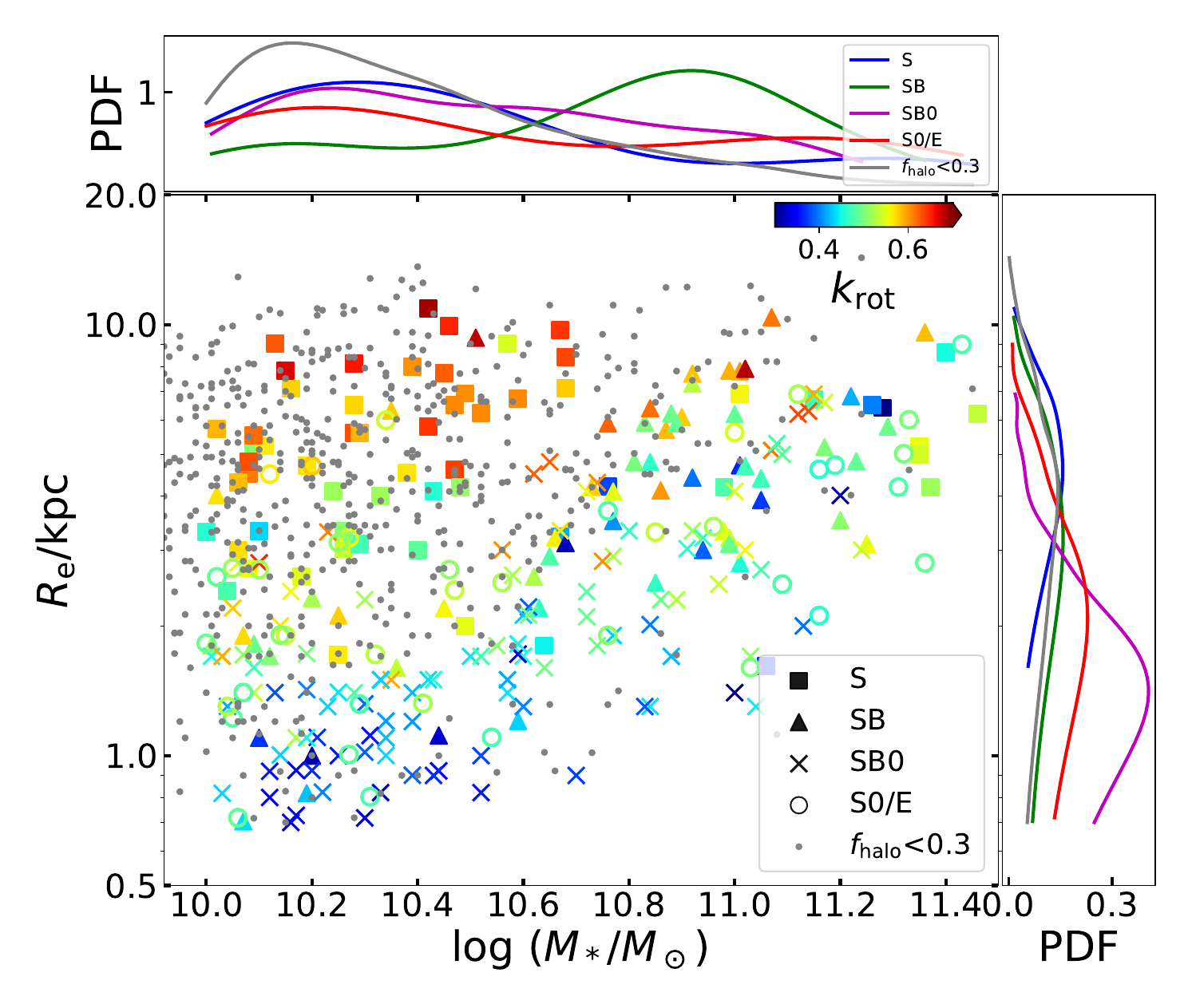}
    \caption{Distribution of Sombrero-like galaxies on the mass-size diagram at \(z = 0\). We conduct a visual classification of such galaxies in spiral (S, squares), barred spiral (SB, triangles), barred lenticular (SB0, crosses), and lenticular/elliptical (S0/E, circles) galaxies. The colorbar represents \(k_\mathrm{rot}\). Gray points indicate other disk galaxies in TNG50 with \(f_\mathrm{halo} < 0.3\).
}
    \label{halfr_M}
\end{figure}

\begin{figure}
	\includegraphics[width=0.9\linewidth]{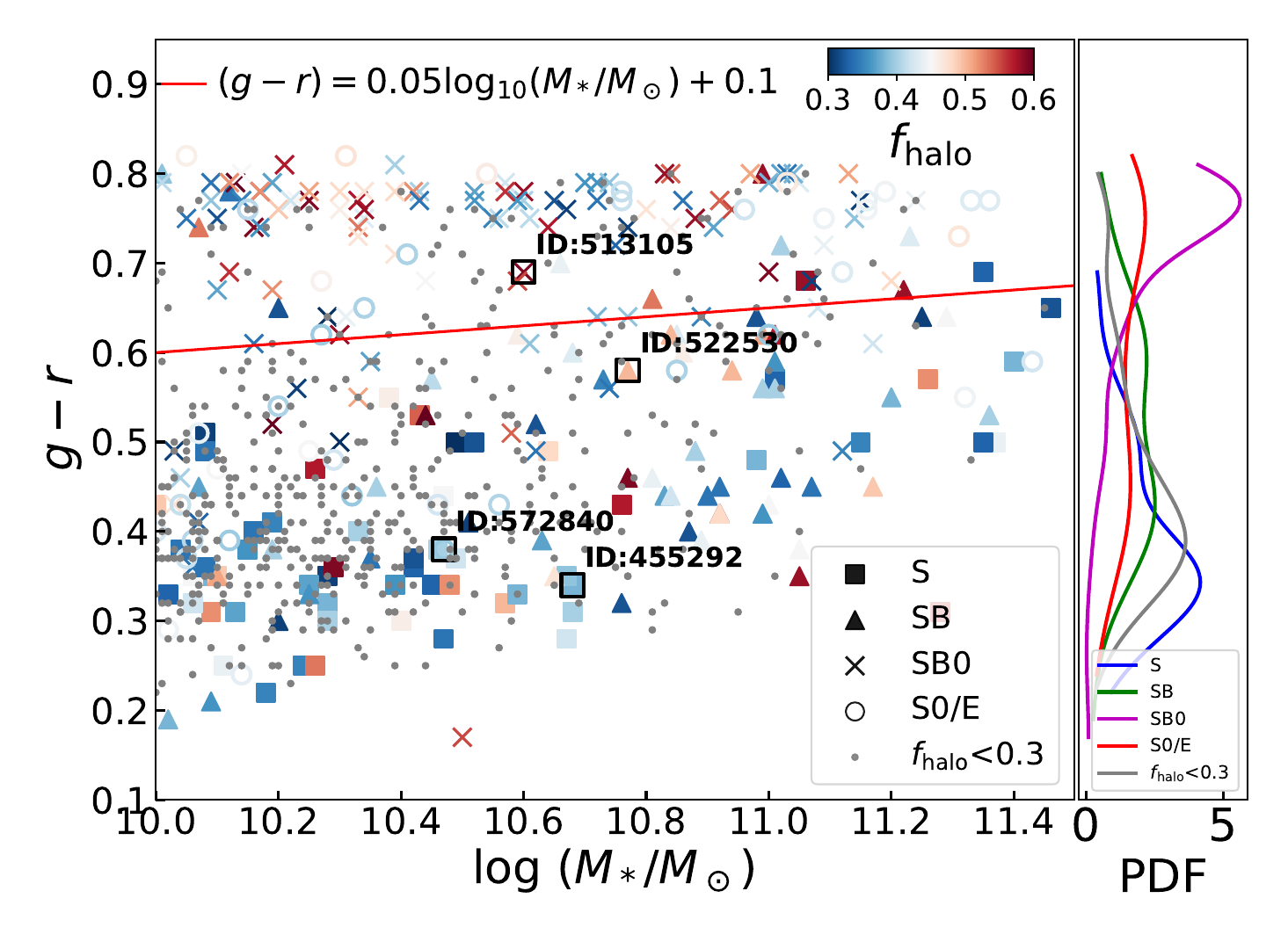}
    \caption{Relation between g-r color and stellar mass for the selected sample of galaxies. The data is based on stellar luminosities from the IllustrisTNG simulations, with g and r magnitudes calculated in the AB system  \citep[see details in][]{2002AJ....123..485S}. The red line ($g-r = 0.05 \log_{10}( M_\star /M_\odot) + 0.1$ mag) serves as the division between early-type and late-type galaxies.
}
    \label{gr}
\end{figure}
    
\section{Diverse morphology, color, and rotation of Sombrero-like, halo-embedded disk galaxies}\label{mcr}

    Figure \ref{example} shows that bars and spirals commonly exist in Sombrero-like galaxies except for those in the leftmost panels. Bars and spirals are generally considered as a strong indication of disks. Consequently, galaxies with such structures are usually classified as rotation-dominated disk and spiral galaxies in the Hubble sequence \citep{1926ApJ....64..321H}. However, from the kinematic perspective, most Sombrero-like galaxies should be classified as elliptical galaxies or slow rotators except for some galaxies with rather large sizes. This is because they possess a large fraction of spheroidal component given by $f_{\rm bulge}+f_{\rm halo}>0.5$ and $\kappa_{\rm rot}<0.5$. Nevertheless, a significant misjudgment takes place when only morphology is taken into account. 

    Here, we conduct a visual classification of the selected Sombrero-like galaxies in their face-on view. The result is shown in Figure \ref{halfr_M}. Most Sombrero-like galaxies have spirals and/or bars, and are thus classified as spiral (S, squares) or barred (SB and SB0, triangles and crosses) galaxies. Only 14 percent of Sombrero-like galaxies are classified as S0/E (circles) galaxies in terms of morphology. The normalized probability distribution is displayed in the side panels. The color of each symbol represents $\kappa_{\rm rot}$, which measures the relative importance of rotation. $\kappa_{\rm rot}>0.5$ has been widely adopted as the criterion for selecting disk galaxies in kinematics. Thus, the morphology-based classification has indeed a systematic difference from the kinematic-based classification. Compact \Somb\ are more likely to be consistently classified as S0/E galaxies by both kinematical and morphological methods due to the lack of spirals and weak rotation. More extended galaxies usually have spirals and stronger rotation and are therefore classified as (barred) spiral galaxies (triangles and squares). Numerical simulations suggested that spiral arms form when the disk is sufficiently dynamically cold \citep[e.g.,][]{1964ApJ...139.1217T,2014RvMP...86....1S}. The morphology of galaxies is mainly determined by the presence of a dynamically cold disk, regardless of whether a massive stellar halo or spheroid exists or not. The existence of a massive stellar halo cannot sufficiently alter the morphological classification of galaxies, except in edge-on views. Compact galaxies are likely to have no spirals, perhaps due to their lack of cold gas or sufficient quenching by AGNs \citep{2024A&A...684A..75C}.
    
    Bars are commonly found in Sombrero-like galaxies in the TNG50 simulation. Thus, bars are not good indicators of the dominance of disk structures in such galaxies either. As suggested by \citet{Lu2024}, mergers play an important role in destroying or preventing bars in massive disk galaxies with $M_*>10^{10.6}M_\odot$, thereby explaining well the presence of unbarred disk galaxies. However, 51 percent of the Sombrero-like galaxies we selected are not included in their sample of disk galaxies selected by $\kappa_{\rm rot}>0.5$. It suggests that many bars can survive after mergers that build up even a massive stellar halo. The existence of massive stellar halos is not enough to suppress the dynamical instability for creating or maintaining bars and spirals in Sombrero-like galaxies. A massive disk is also not necessary for forming or maintaining bars. Furthermore, bars are likely to be present in compact galaxies even if their overall rotation is weak. This is consistent with the result of \citet{Lu2024}, which shows that a compact morphology promotes the formation of bars in disk galaxies. Furthermore, it is important to stress that we are not suggesting bars and spirals are insignificant structures in galaxies. Rather, we highlight that traditional morphological classification overlooks the stellar halo, which serves as a crucial indicator of the merger history a galaxy has undergone. However, it is a long history that bulges are typically considered the primary indicator of mergers.
    
    In galaxies resembling the Sombrero, the disk structures are typically formed through continuous star formation following a merger event, as proposed by \citet{2021ApJ...919..135D}. Consequently, it is not unexpected that many of them still have active star formation up to $z = 0$. In Figure \ref{gr}, we present the color-mass diagram of the Sombrero-like galaxies. We use the data of the $r$-band and $g$-band from the IllustrisTNG public data release without considering the effect of dust extinction. The depicted red line represents the equation $(g - r)=0.05\log_{10}(M_\star/M_\odot)+0.1mag$ \citep{2020A&A...641A..60P}, which is widely employed as a boundary between early-type (red sequence) and late-type (blue cloud) galaxies. It is evident that Sombrero-like galaxies have a wide distribution in the color-mass diagram. There is no tendency for Sombrero-like galaxies to be either star-forming or quiescent. In other words, a more violent merger in the past does not result in a significant quenching of star formation in Sombrero-like galaxies. Indeed, there are numerous galaxies in which the spheroidal components, indicated by $f_{\rm halo}+f_{\rm bulge}>0.5$, are dominant, yet they display remarkably blue colors. Sombrero-like galaxies, because of their dual nature - being either star-forming or quiescent - pose a challenge for identification based solely on color. Their color does not offer a clear classification, making it difficult to categorically define them as early-type or late-type galaxies. 

    In conclusion, classifying galaxies by either morphological features (like bars and spirals) or color is likely to involve a high degree of uncertainty in their intrinsic stellar halo structures. These halo-embedded disk galaxies exhibit diverse disk properties while maintaining morphological similarities to the Sombrero galaxy. Thus, a large uncertainty exists in their merger histories. A blue color and disk character are not enough to identify disk galaxies robustly. There is no effective method that can readily distinguish Sombrero-like galaxies from those dominated by disk structures in terms of both morphology and color. Thus, their differences in merger history are likely to cause substantial uncertainty in our understanding of the formation and evolution of galaxies. Despite the diverse morphology and color, many \Somb\ have relatively more compact morphology than pure-disk galaxies with $f_{\rm halo}<0.3$ (gray dots in Figure \ref{halfr_M}). 

\section{Comparison between morphological and kinematical decomposition of Sombrero-like, halo-embedded disk galaxies}\label{result}

\begin{figure*}
\centering
	\includegraphics[width=0.7\linewidth]{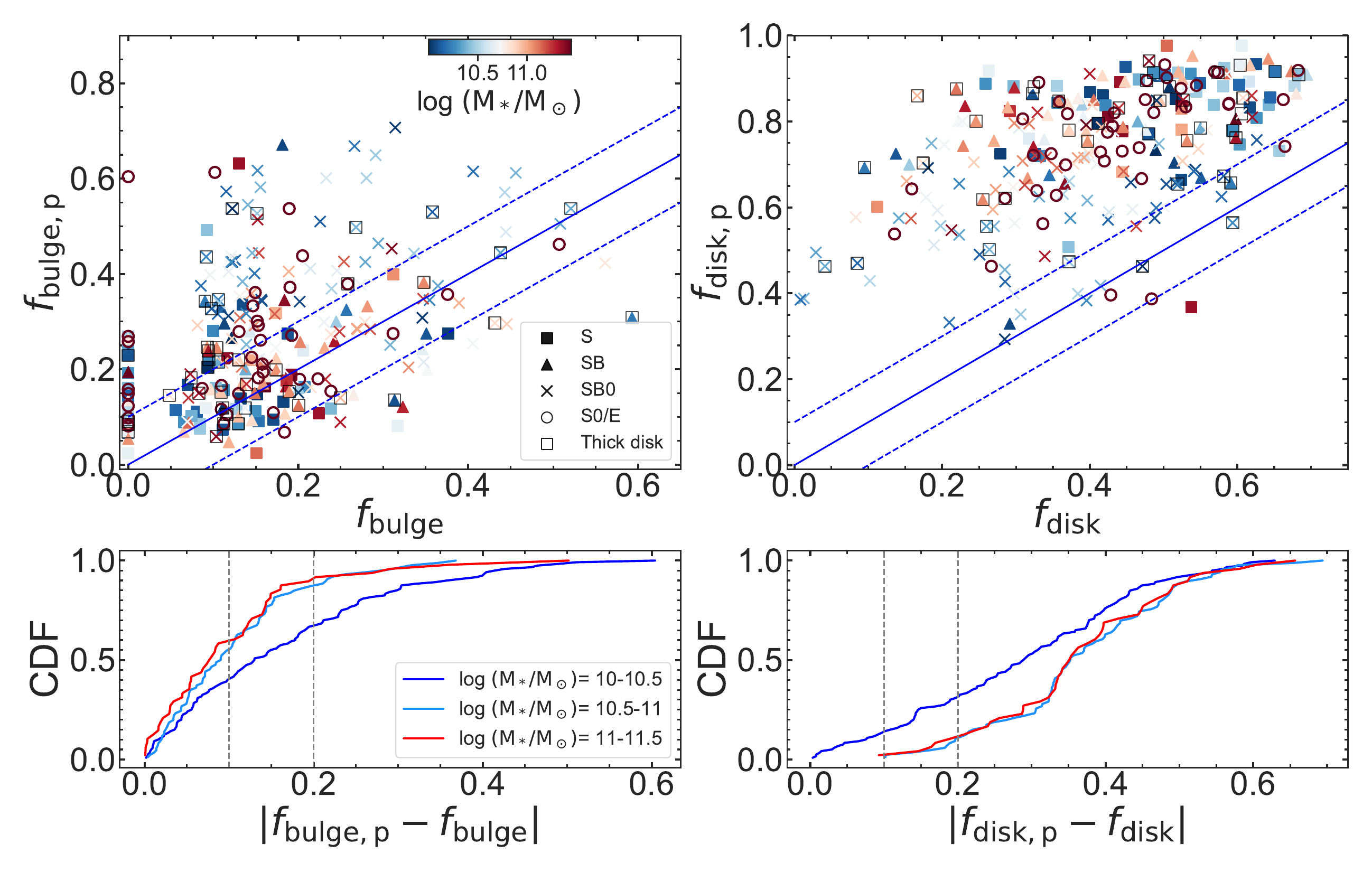}
    \caption{Comparison between the photometric and kinematic decomposition of the Sombrero-like galaxy in face-on views. The top panels show, from left to right, the relationship between the distributions of $f_\text{bulge,p}$ vs. $f_\text{bulge}$ and $f_\text{disk,p}$ vs. $f_\text{disk}$. The different symbols represent the visual classification of each Sombrero-like galaxy. The bottom panels show the accumulative distribution profiles of their relative differences. The galaxies are color-coded by mass, with red indicating high-mass galaxies and blue indicating low-mass galaxies. The blue solid line represents $y = x$, while the dashed blue lines represent the $\pm0.1$ deviation range from $y = x$. Galaxies with a fit stellar halo ellipticity of less than 0.3 are considered thick disks and are highlighted with black rectangles. The galaxies are divided into three mass bins: log$(M_\star/M_\odot) = 10-10.5$ (blue), $10.5-11$ (light blue), and $11-11.5$ (red). Additionally, instances where $f_{\rm bulge} = 0$ suggest that no bulge is sufficiently prominent to be detected by the kinematic decomposition method, as noted by \citet{2021ApJ...919..135D}. The gap appearing around $f_{\rm bulge} \sim 0.03$ lacks physical significance. Furthermore, the morphological decomposition method inherently identifies any excess luminous structure as a bulge, regardless of whether it exhibits significant rotation or not.
}
    \label{GAfaceon}
  \end{figure*}

\begin{figure*}
    \centering
	\includegraphics[width=0.9\linewidth]{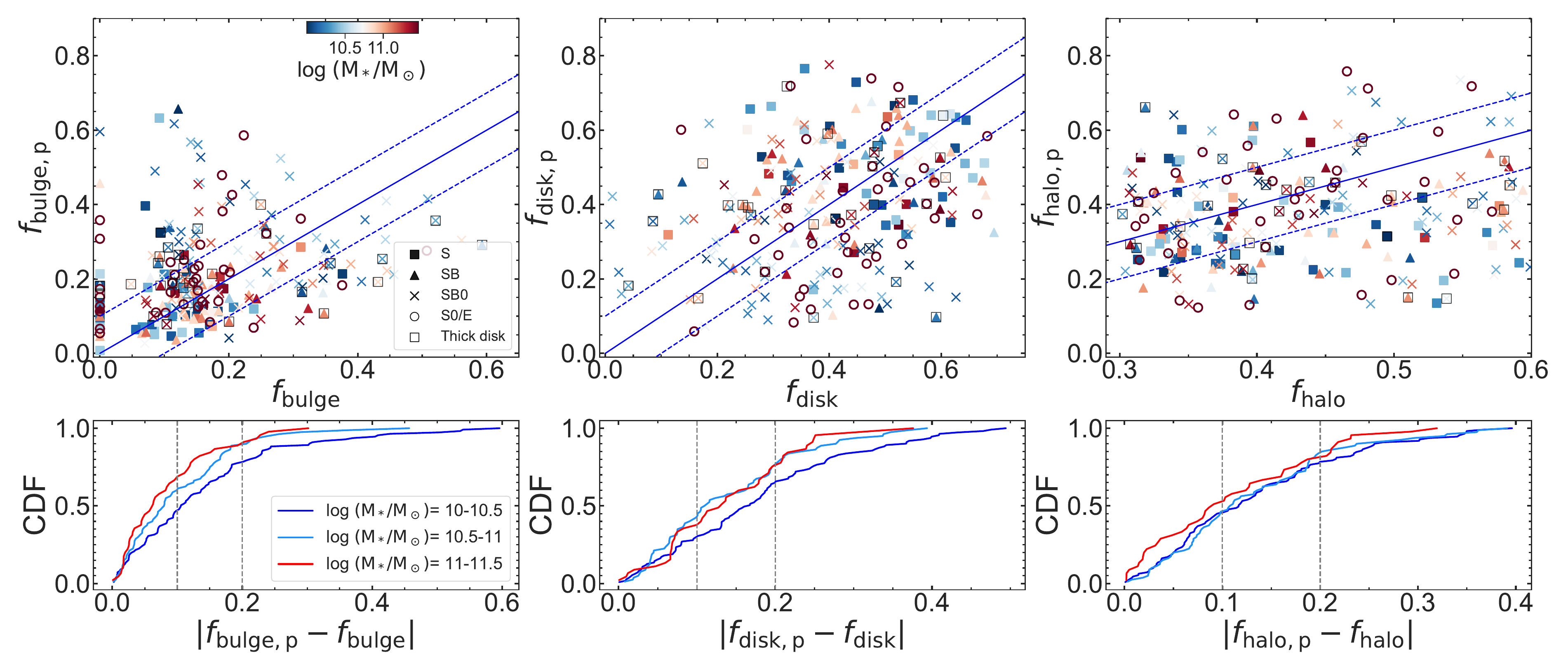}
    \caption{Comparison between the photometric and kinematic decomposition structures of Sombrero-like galaxies in edge-on views. The top panels show, from left to right, the relationship between the distributions of $f_\text{bulge,p}$ vs. $f_\text{bulge}$, $f_\text{disk,p}$ vs. $f_\text{disk}$, and $f_\text{halo,p}$ vs. $f_\text{halo}$. This figure uses the same convention as \reffig{GAfaceon}. 
}

    \label{GAedgeon}
\end{figure*}

    Morphological decomposition is one of the most important ways to classify galaxies and understand their underlying physics. Different galaxy structures emit overlapping light, which makes it a difficult task to accurately decompose, especially the stellar halo in Sombrero-like galaxies. We further use the mass ratio of kinematic structures measured by {\tt auto-GMM} from \citet{2019ApJ...884..129D} and \citet{2020ApJ...895..139D}. Generally, galaxies are classified into cold/warm disk, bulge, and stellar halo structures according to the binding energy and angular momentum of their stars. \citet{2021ApJ...919..135D} offers strong evidence that different structures have not only distinguishable kinematic properties but also different formation mechanisms. As shown in Section \ref{mcr}, \Somb\ have various properties regarding structures, color, and even rotation. In this section, we will compare the results of photometric decomposition and kinematic decomposition for selected galaxies and evaluate the effectiveness of morphological decomposition.

\begin{figure}
	\centering
    \includegraphics[width=0.9 \linewidth]{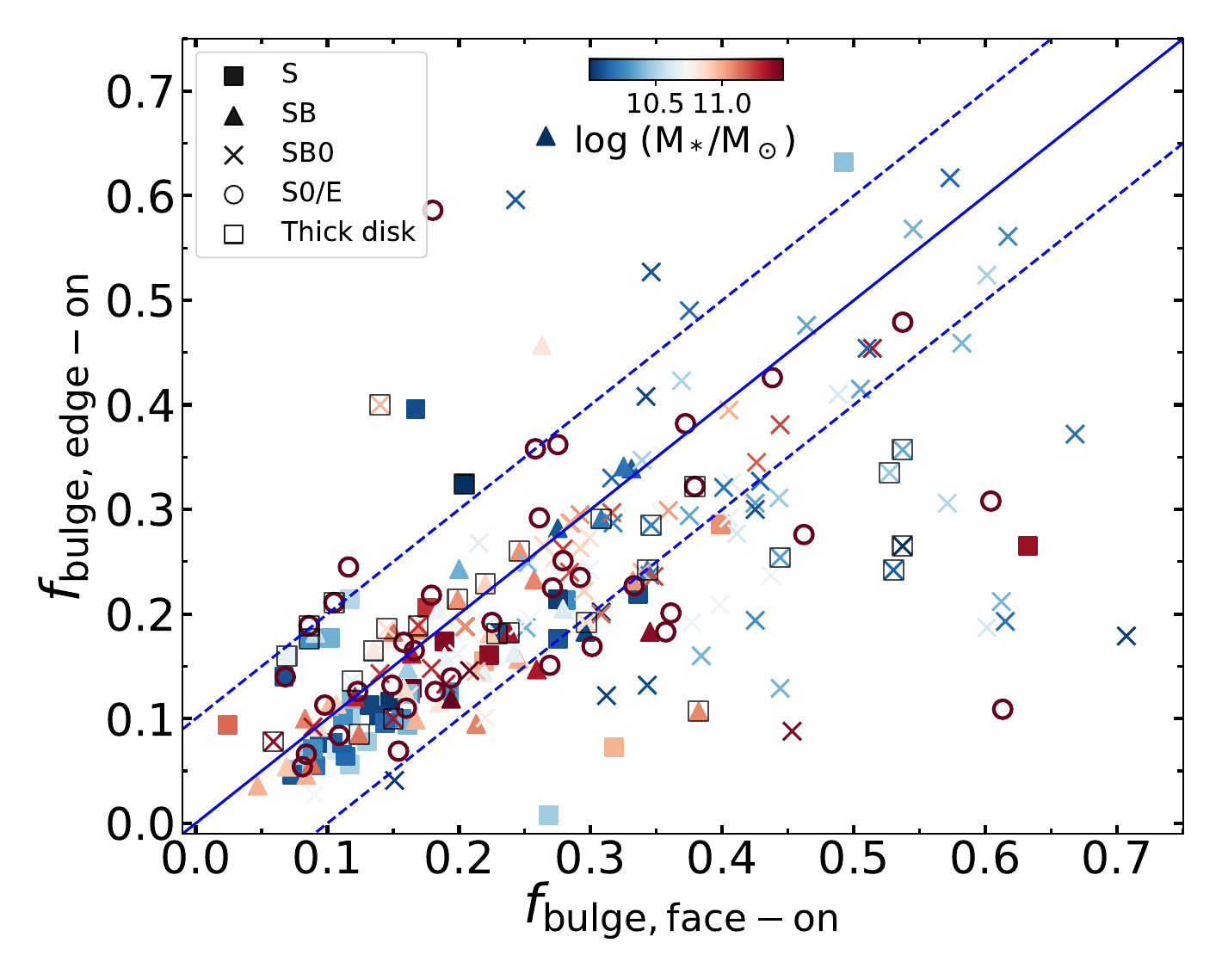}
    \caption{Relationship between the bulge-to-total ratio derived from GALFIT two-component decomposition (bulge+disk) in face-on galaxies and the three-component decomposition (bulge+disk+stellar halo) in edge-on galaxies. This figure uses the same convention as Figure~\ref{GAfaceon}.}
    \label{BTfe}
\end{figure}

\subsection{Bulge+disk decomposition at a low inclination angle: The disk mass fraction is significantly overestimated in morphological decomposition.}\label{sec:faceon}

    In Sombrero-like galaxies with low or moderate inclinations, stellar halos are generally not distinguishable (for example, see the upper two rows in Figure \ref{example}). Although a bulge+disk decomposition can fit their surface densities reasonably well, the mass fraction of disk structures is usually overestimated. It therefore induces a large difference between the results of photometric and kinematic decomposition. In face-on views, $f_\text{disk,p}$ is typically overestimated by at least 0.2, as shown in the right panel of Figure \ref{GAfaceon}. This is due to the fact that the massive stellar halos of Sombrero-like galaxies contribute significantly to the disks in terms of morphology. This difference can be even greater when bars are regarded as part of the disks. Such a difference has a weak dependence on their stellar masses, as shown in three mass ranges of $\log(M_\star / M_\odot)=10-10.5$ (blue), $10.5-11$ (light blue), and $11-11.5$ (red).

    Although the bulge+disk morphological decomposition is unsuccessful in measuring disk structures, it usually measures bulges okay, as shown in \reffig{GAfaceon}. Most galaxies (more than 70\%) have small bulges with $f_{\rm bulge}<0.2$ as measured by both morphological and kinematical methods, even though massive stellar halos exist in such Sombrero-like galaxies. It is clear that bulges are overestimated in barred galaxies (shown by crosses). This is understandable because bars are commonly misclassified as bulges in the bulge+disk decomposition. Accurately decomposing bulges from bars is a difficult task. It is even more difficult in TNG50 because, as reported in \citet{Lu2024}, there are a large number of galaxies with short bars. In this study, we do not attempt to decompose bars, and this will not influence our results related to stellar halos. Our result here challenges the common view that bulges measured in morphology are regarded as a good indicator of mergers.
    
\subsection{Bulge+disk+stellar halo decomposition in edge-on views: reasonably well in morphology but large uncertainty}\label{sec:edgeon}

    The bulge+disk+stellar halo decomposition is carried out for the edge-on views of the same sample of Sombrero-like galaxies. In Figure \ref{GAedgeon}, we compare the mass fractions of bulges, disks, and stellar halos (from left to right) with the results of kinematic decomposition for Sombrero-like galaxies. As shown in the first row, most galaxies have $f_{\rm bulge,p}<0.2$, though there is a large scatter. This result is consistent with that of the bulge+disk decomposition in face-on views. It should be emphasized that many barred galaxies (marked by crosses) also lead to massive bulges when measured morphologically, because there is no adequate method to separate bars from bulges, especially in edge-on views. As can be seen in \reffig{BTfe}, the bulges measured in face-on views (on the $x$ axis) and edge-on views (on the $y$ axis) are roughly similar, except for some barred galaxies. The face-on measurement we use here further overestimates the mass fraction of bulges without including a bar component. {\tt auto-GMM} also cannot accurately divide bars from other structures in kinematics \citep{2019ApJ...884..129D}. In conclusion, the bulges in the face-on bulge+disk decomposition are likely to match those in the edge-on bulge+disk+halo decomposition in terms of morphology, and both of them are somehow similar to the kinematically defined bulges. Despite the unrealistic low S\'ersic index of galaxies in TNG simulations, a rather small bulge is obtained not only through kinematic but also morphological decomposition. This result is in agreement with the finding of \citet{2012MNRAS.423..877G}, although a statistical analysis is still needed. Both morphological and kinematical measurement shows that most \Somb\ generally have low-mass bulges with $f_{\rm bulge}<0.2$. The existence of galaxies with bulge mass ratio $>0.3$ is largely due to their bars. 

    Morphologically defined stellar halos are positively correlated with those defined by kinematic methods, though there is a rather large scatter, shown in the right panel of \reffig{GAedgeon}. Nearly 50\% of galaxies can be well estimated by photometric decomposition with a difference of less than 0.1, while the structures in about 70\% of galaxies are properly estimated if a difference of 0.2 is acceptable. It is worth noting that stellar halos are generally underestimated using photometric decomposition when $f_{\rm halo}>0.45$, even in edge-on views. The extended envelope measured in edge-on views seems to be a good method for accurately identifying and decomposing stellar halos, so as to further quantify the strength of mergers that each galaxy has undergone. 
    
    In summary, our results suggest that face-on bulge-disk decomposition can roughly identify bulges. The bulge+disk+stellar-halo decomposition in the case of nearly edge-on galaxies is required to better measure stellar halos. \citet{2020ApJ...895..139D} suggested that kinematically derived structures are systematically different from morphologically defined bulges and disks. This is mainly because of the degeneracy of structures in morphology, especially the poor measurement of stellar halos. Furthermore, the use of the S\'ersic function, which has no clear physical meaning as discussed in Section \ref{sec:sersic}, would also cause a large deviation between structures defined by morphological and kinematical methods.

\subsection{Difference between the mass ratio of stellar halos at different radii measured by kinematic and morphological decompositions}

\begin{figure*}
    \centering
	\includegraphics[width=0.8\linewidth]{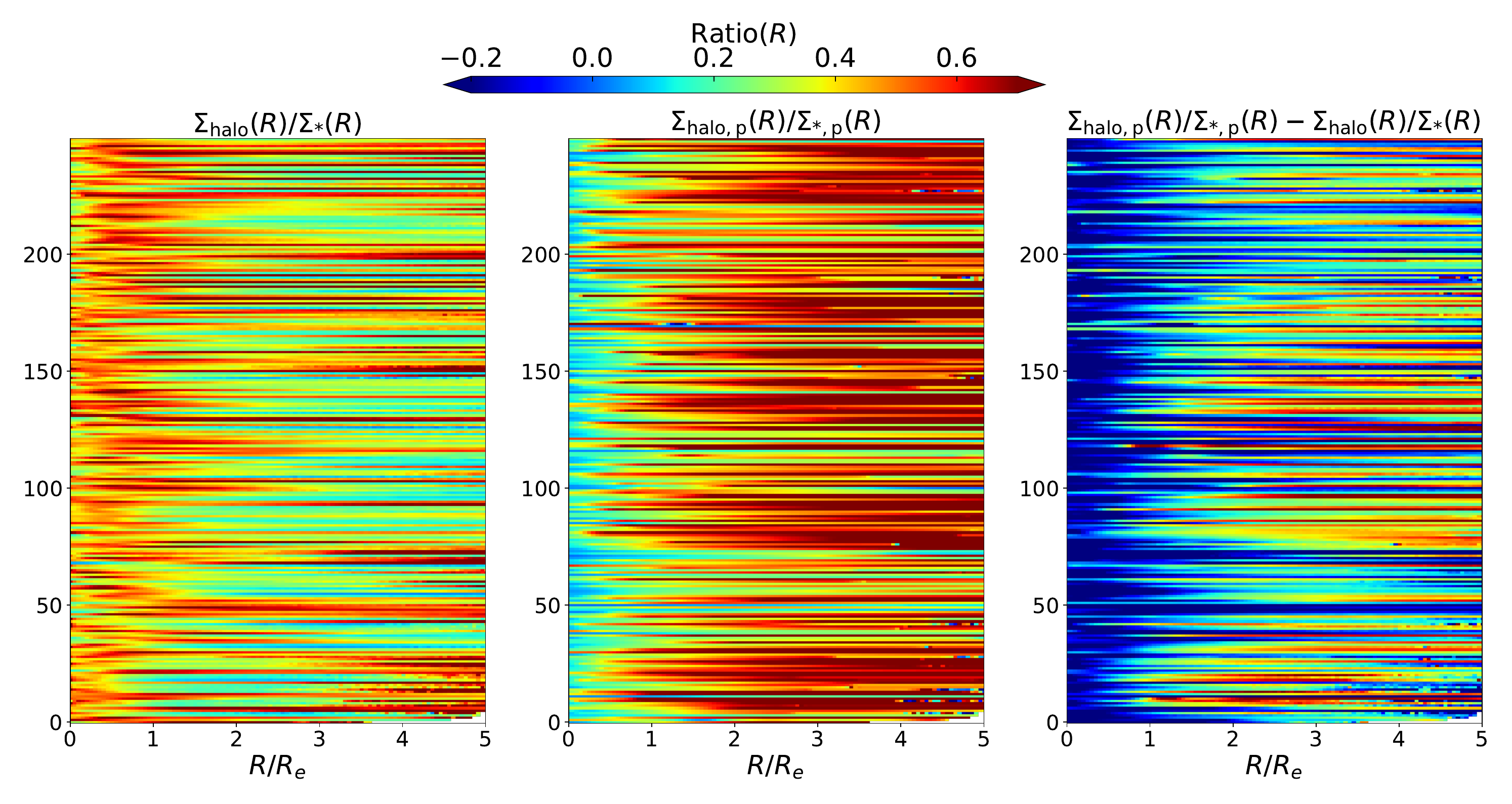}
    \caption{The relationship between the kinematically derived stellar halo mass fraction, $\Sigma_{\text{halo}}(R) / \Sigma_{*}(R)$ (left panel), the photometrically derived stellar halo luminosity fraction, $\Sigma_{\text{halo,p}}(R) / \Sigma_{*,\text{p}}(R)$ (middle panel), and their difference (right panel) as a function of the radius within $5 R_e$. Each row represents an individual galaxy. 
}
    \label{halo_profile}
\end{figure*} 

Although the mass ratio of the stellar halo obtained from morphological analysis is generally consistent with that derived from kinematic decomposition, their radial distributions show a systematic difference, even when measured in edge-on views. \reffig{halo_profile} presents the stellar halo fractions obtained through both kinematic (left panel) and morphological (middle panel) decomposition. These fractions are calculated by dividing each major axis slice's stellar halo mass (or luminosity) by the corresponding total stellar mass (or luminosity) within that slice in edge-on views. The right column shows the difference between them. All results are plotted as functions of radius within 5\(R_e\). 
\begin{table}[htbp]
\caption{Detection probability of stellar halos at different redshifts.
}
\centering
\renewcommand{\arraystretch}{1.6} 
\scalebox{0.6}{ 
\large
\begin{tabular}{@{\hskip 1mm}l|ccccc} 
\toprule
\diagbox[width=3cm]{log$(M_\star/M_\odot)$}{Z} & $z=0.091$ & $z=0.2$  & $z=0.3$  & $z=0.4$  & $z=0.5$    \\ 
\midrule
$10 - 10.5$ & $0.0296^{+0.0002}_{-0.0045}$ & $0.0292^{+0.0005}_{-0.0048}$ & $0.0292^{+0.0005}_{-0.0063}$ & $0.0289^{+0.0008}_{-0.0074}$ & $0.0286^{+0.0011}_{-0.0088}$  \\ 
$10.5 - 11$ & $0.0601^{+0.0008}_{-0.0033}$ & $0.0599^{+0.0009}_{-0.0058}$ & $0.0598^{+0.0009}_{-0.0075}$ & $0.0595^{+0.0007}_{-0.0098}$ & $0.0594^{+0.0006}_{-0.0130}$  \\ 
$11 - 11.5$ & $0.1009^{+0.0011}_{-0.0065}$ & $0.1003^{+0.0014}_{-0.0085}$ & $0.0996^{+0.0018}_{-0.0110}$ & $0.0989^{+0.0024}_{-0.0129}$ & $0.0987^{+0.0021}_{-0.0155}$  \\
\bottomrule 
\end{tabular}}
\label{probability}
\tablefoot{Surface brightness detection limit of 28 mag is used. The positive deviation assumes a limit of 30 mag and the negative deviation assumes 26 mag. The central region (within 3 kpc radius) is masked to isolate the diffuse envelope, representing the stellar halo.}
\end{table}

It is clear that the ratio obtained from the kinematic method, \(\Sigma_{\text{halo}}(R) / \Sigma_{*}(R)\), is significantly higher than that from the morphological method $\Sigma_{\text{halo,p}}(R) / \Sigma_{*,p}(R)$ within \(R < 1.5R_e\). The right column of the figure highlights the differences between the two methods: within \(1.5R_e\), the kinematic mass distribution is approximately 0.3 higher than the photometric result. This discrepancy is largely attributed to the exponential assumption used in morphological decomposition, whereas both simulations \citep{2020ApJ...895..139D} and IFU observations \citep{2018MNRAS.473.3000Z} suggest that central truncations of disk orbits are common in galaxies. This leads to a substantial underestimation of stellar halos and an overestimation of disks in the central regions of galaxies. 

On the other hand, the stellar halo mass ratio obtained from photometric decomposition tends to be higher than that from kinematic decomposition beyond \(1.5R_e\). Beyond this radius, the kinematic mass distribution is approximately 0.2 lower than the photometric result. This is likely because the stellar halo is composed of older stars, making it brighter in the \(r\)-band, while dust extinction has a more pronounced effect on edge-on disks. Furthermore, the coupling effect in the outer regions is weaker, meaning the contribution of the halo component becomes more dominant.

\section{Discussions}\label{sec:discution}
\subsection{Stellar halos as probes of galactic merger histories}\label{Exsitu}
 
\begin{figure}
    \includegraphics[width=1
    \linewidth]{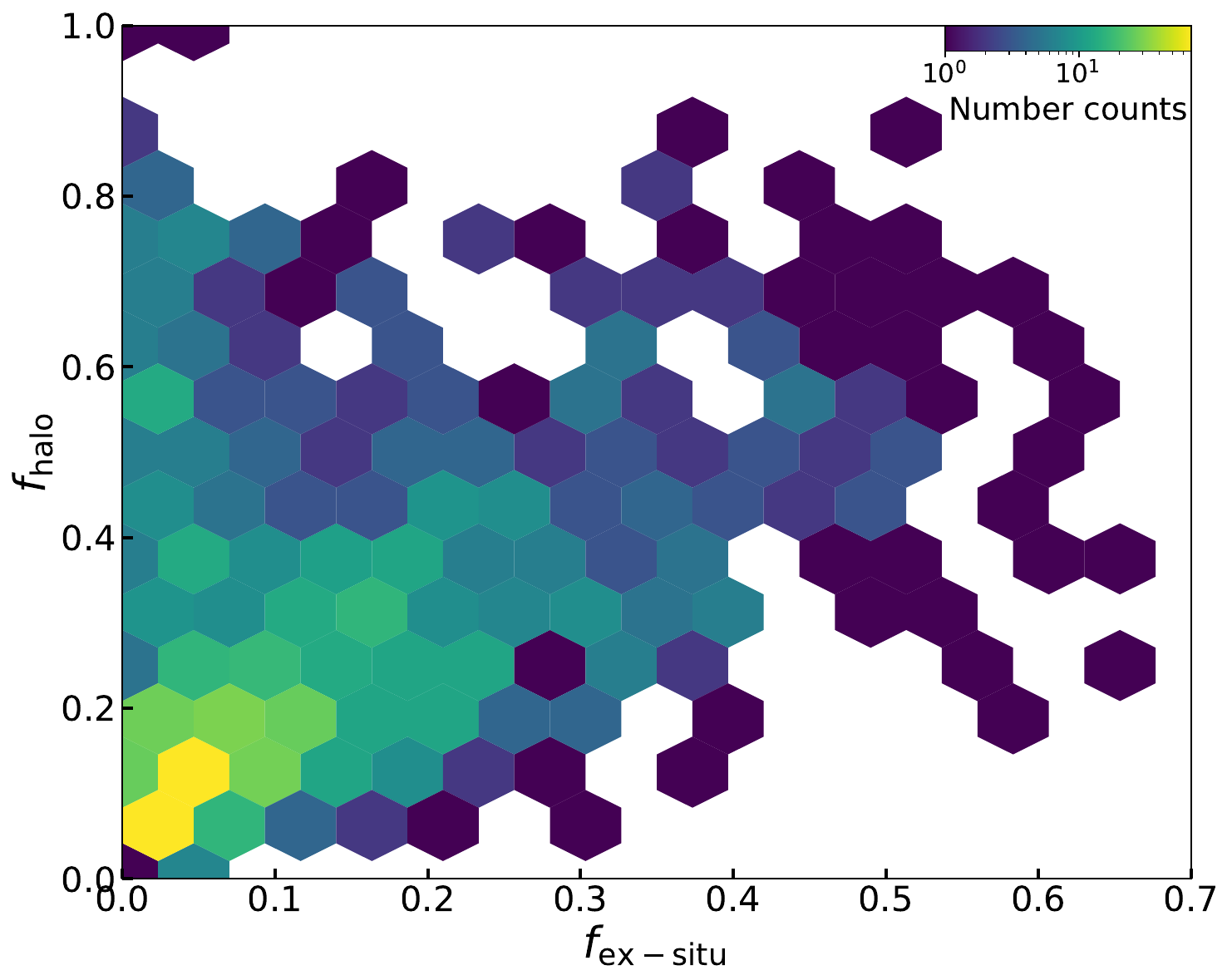}
    \caption{Distribution of TNG50 galaxies with stellar mass $M_\ast > 10^{10}\ M_\odot$ and $0.3 < f_\mathrm{halo} < 0.6$ on the $f_\mathrm{ex\text{-}situ}$ versus $f_\mathrm{halo}$ diagram. The color indicates the number of galaxies in each bin.
}
    \label{exsitu}
\end{figure}

Stellar halos are pivotal tracers of galactic assembly processes, primarily composed of ex-situ stars accreted through hierarchical mergers over cosmic time \citep[e.g.,][]{2014MNRAS.444..237P,2005ApJ...635..931B}. The ex-situ stellar mass fraction (\(f_\mathrm{ex-situ}\)) exhibits a strong positive correlation with \(f_\mathrm{halo}\), as illustrated in Figure~\ref{exsitu}. Here, \(f_\mathrm{ex-situ}\), measured following the methodology of \citet{2016MNRAS.458.2371R}, quantifies the fraction of stars accreted from external sources during hierarchical mergers. This correlation reinforces the hypothesis that stellar halos predominantly form through ex-situ accretion of disrupted satellite galaxies, making them effective tracers of stellar populations of external origin. It is worth mentioning that \(f_\mathrm{halo}\) is typically larger than \(f_\mathrm{ex-situ}\) because \(f_\mathrm{halo}\) also includes stars originating from the in-situ population of the progenitor galaxy. Moreover, tidal interactions or harassment processes—such as high-speed encounters in dense environments—can also elevate \(f_\mathrm{halo}\) without significantly increasing \(f_\mathrm{ex-situ}\), complicating direct interpretations of merger histories. To disentangle these effects, multiwavelength photometric studies combined with detailed kinematic measurements are essential. Such approaches enable the characterization of accretion events and differentiate between merger-driven halo growth and secular processes like tidal stripping. This result is qualitatively consistent with other analyses using, for example, Illustris \citep{2014MNRAS.444..237P} and Auriga \citep{2019MNRAS.485.2589M}, showing that the outer regions of stellar halos are from disrupted satellite galaxies. Thus, stellar halos can serve as probes of galactic merger histories, forming the basis for machine learning predictions \citep{Cai2025}, though the associated uncertainties remain substantial.

\subsection{Prediction of the probability of the occurrence of a Sombrero-like galaxy at different redshifts}\label{Prediction}

    We shift the surface luminosity of galaxies to redshifts varying between 0.1 and 0.5 to estimate the probability of a Sombrero-like galaxy’s appearance measured in $r$-band. The surface brightness decreases as a function of redshifts by \( (1 + z)^{-4} \) \citep{1930PNAS...16..511T} due to the increase in distance and the cosmological expansion. Both the observed central wavelength \( \lambda \) and bandwidth \( \Delta \lambda \) increase by a factor of \( (1 + z) \). Then we estimate the surface brightness decrease proportional to \( \propto (1 + z)^{-3} \). Here we ignore the detail changes of Sombrero-like galaxies during these time. A Sombrero-like galaxy can be identified when its surface luminosity at a large radius (here, > 3 kpc) exceeds the limiting luminosity of a certain telescope. Furthermore, as indicated in Section \ref{result}, it is very hard to identify stellar halos in cases with low or moderate inclination angles. Thus, the probability of detecting Sombrero-like galaxies is approximately 22\%, provided the inclination angle \( i > 70^\circ \). The probability of the occurrence of a Sombrero-like galaxy is then approximated by 22\% times the probability of Sombrero-like galaxies in the TNG50 simulation for a given redshift.

    As shown in Table \ref{probability}, the probability is estimated under a typical limiting magnitude of 28 mag for the HST and future CSST space telescopes. Here, we also present the predicted detection probabilities that deviate from the 28 mag cases for a limiting magnitude of 26 mag (lower values) and 30 mag (upper values). The \Somb\ are further separated into three mass ranges.Clearly, a more massive Sombrero galaxy is more easily identified, with a probability of 0.06 when log \((M_\star/M_\odot)>10.5\) and approximately 0.03 when log \(10<(M_\star/M_\odot)<10.5\). It is important to note that our results represent the upper limit of probability in a scenario that approximates realism, as we have not accounted for the evolution of \Somb\ or the difficulties associated with subtracting background contamination and noise.

\subsection{Puzzles about the physical meaning of S\'ersic index of bulges}\label{sec:sersic}

\begin{figure}
	\includegraphics[width=0.9\linewidth]{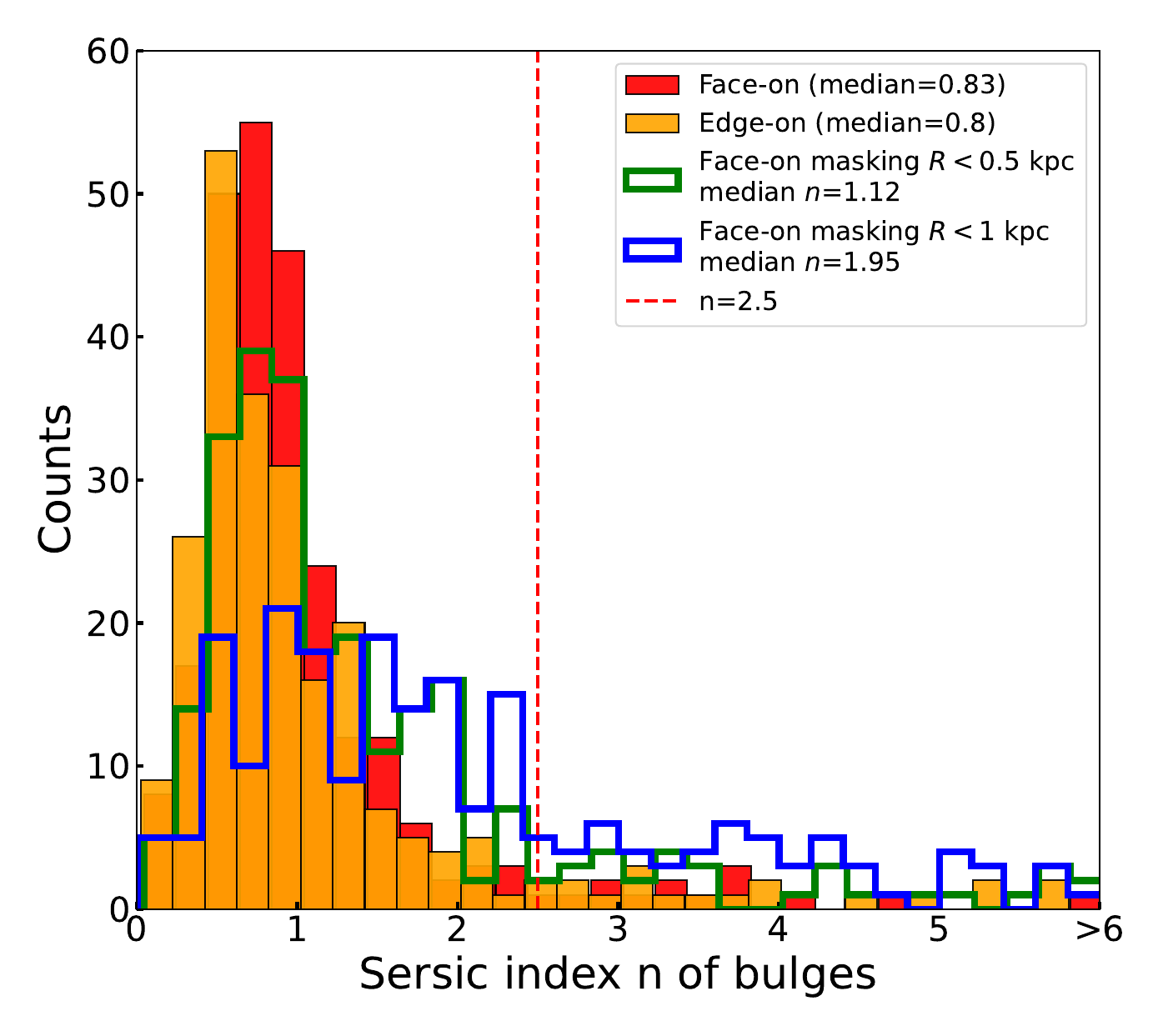}
    \caption{Statistics of the S\'ersic index obtained from {\tt GALFIT} fitting for galaxies in both face-on and edge-on orientations, as well as for cases with masking applied to the galaxy center at 0.5 kpc and 1 kpc, are shown. The dashed red line indicates a S\'ersic index of $n=2.5$. The majority of the galaxies have a S\'ersic index less than 2.5.
}
    \label{sersicn}
\end{figure}

    Classical bulges have long been considered as an output of mergers. The Sombrero galaxy is generally regarded as a typical example whose disk is embedded in a large and massive classical bulge. The S\'ersic index, \(n\), has long been used as a reliable indicator for differentiating between different types of bulges \citep{2004ARA&A..42..603K,2008AJ....136..773F,2008IAUS..245..117G}. Generally, classical bulges have \(n > 2.5\), and cases with lower \(n\) are classified as pseudo-bulges \citep[e.g.,][]{2008AJ....136..773F,2009ApJS..182..216K,2016ASSL..418...41F}. The co-evolution between classical bulges and their central super-massive black holes is still being actively debated \citep[e.g.,][]{2013ARA&A..51..511K, 2017ApJ...836..181D,  2019MNRAS.490.3196P,2020ApJ...888...65G,2021PhyU...64..386N,2022ApJ...941...95W}. However, the physical meaning of the S\'ersic function and index is unclear. Thus, we investigate the S\'ersic index of \Somb\ to explore any possible causal connections with the intrinsic structures of galaxies derived from the kinematic method.

    Figure \ref{sersicn} shows the distribution of the S\'ersic index obtained from {\tt GALFIT} fits for galaxies in both face-on and edge-on views. Most of the Sombrero-like galaxies give \(n < 2.5\). The median values of \(n\) are 1.11 and 1.24 when measured in face-on and edge-on views, respectively. From an observational perspective, most bulges in \Somb\ are thus pseudo-bulges. This result is in line with \citet{2019MNRAS.483.4140R} and \citet{2020ApJ...895..139D}, which states that the TNG simulations have difficulty in reproducing the high S\'ersic indices of galaxies and classical bulges. It indicates that classical bulges have no clear distinction of $n$ from pseudo-bulges. 

    A distribution having a large S\'ersic index demands a large central peak and an outer envelope that curves upward in the logarithmic space with respect to a simple exponential profile. \Somb\ are an excellent sample for exploring the physical significance of the S\'ersic function. As proposed by \citet{2020ApJ...895..139D}, there is no single kinematically-defined structure that corresponds to the properties of classical bulges as defined in observations. Consequently, a classical bulge might be a composite structure. Kinematic bulges, which are characterized by low binding energy and are thus tightly bound in the central regions of galaxies, have generally small size. As a result, they contribute to the central part of the S\'ersic function. Additionally, the outer part might correspond to either kinematically-defined stellar halos, disks, or their twists from an exponential profile.

    Galaxies in TNG simulations may not develop high enough central concentrations to reproduce galaxies with large S\'ersic index, given its spatial resolution limitations. This may result in a S\'ersic index that is lower than what is observed in real galaxies. We thus mask the central regions of face-on galaxies at $R<0.5$ kpc and $R<1$ kpc when the {\tt GALFIT} fitting is performed to reduce any potential numerical problems in the central regions. The results of masking \(R < 0.5\) kpc (magenta) and \(R < 1\) kpc (green) are also presented in Figure \ref{faceon}. The S\'ersic index of the galaxy in the second row (dashed green) becomes considerably larger, while there is no obvious change in their overall fitting results. As shown in Figure \ref{sersicn}, the S\'ersic index statistically increases as the masked region increases. There are far more bulges having \(n>2.5\) in the case of masking regions with \(R > 1\) kpc. 
    
\begin{figure}
	\includegraphics[width=0.9 \linewidth]{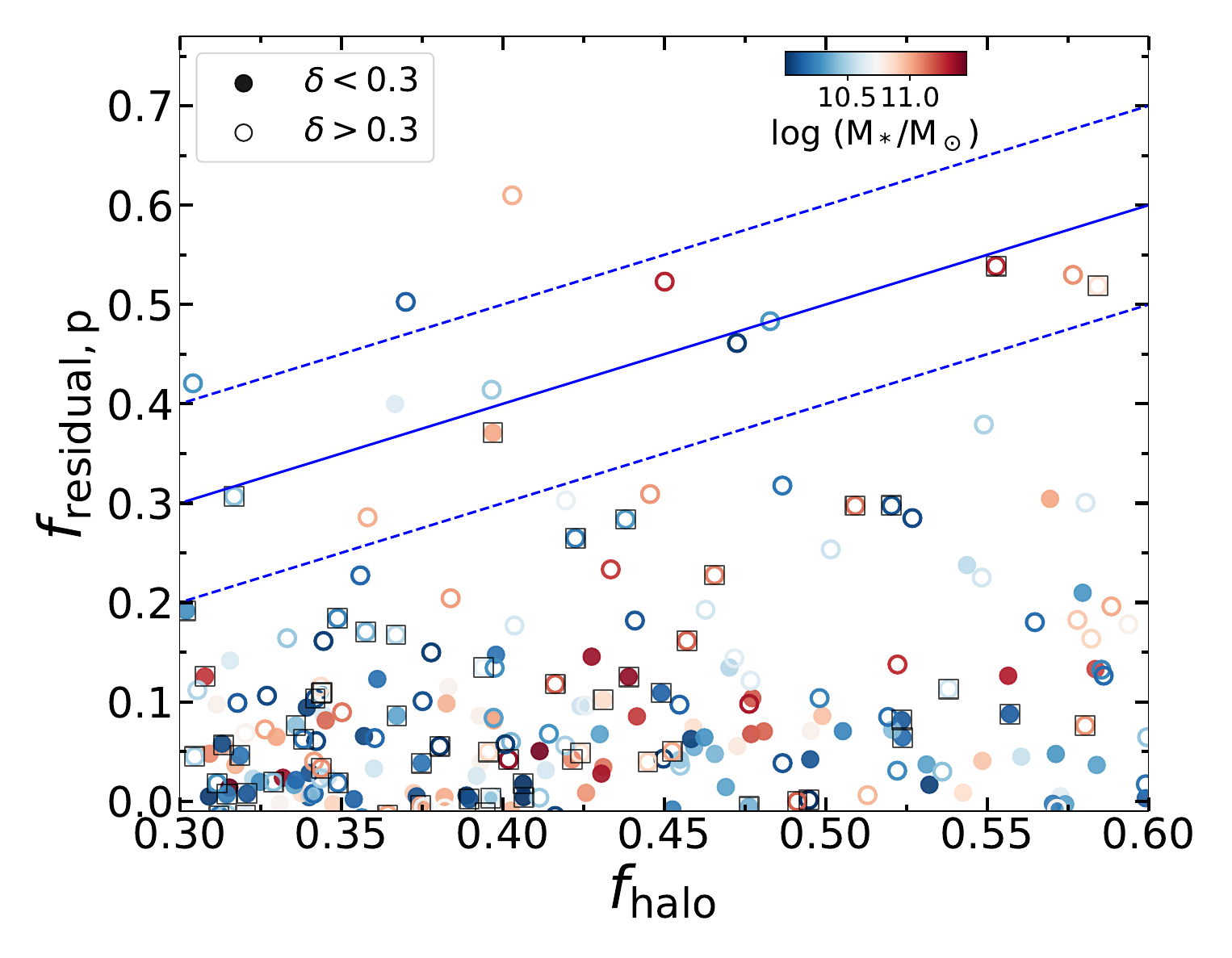}
    \caption{Relationship between the residual \( f_{\text{residual, p}} = \frac{L_{\text{mock}} - L_{\text{model}}}{L_{\text{mock}}} \) from the GALFIT two-component fitting for face-on galaxies and the stellar halo fraction \( f_{\text{halo}} \) derived from kinematic decomposition. Galaxies with relative errors less and larger than 0.3 use solid and open symbols, respectively. The cases marked with black rectangles are galaxies with thick disk structures.
}
    \label{residual}
\end{figure}

    We see no obvious indication that the S\'ersic index is strongly correlated with massive stellar halos as well as the Hubble types. Massive stellar halos may cause a higher surface brightness in the outer regions of galaxies compared to a simple exponential profile, thereby resulting in a larger \(n\). Therefore, the presence of a more massive stellar halo may lead to a poorer fitting result when only two components are used. In this situation, a larger residual \(f_{\rm residual,p}\) should be proportional to \(f_{\rm halo}\). However, once again, we do not see any clear evidence of the existence of stellar halos, shown in Figure \ref{residual}. Consequently, we have not found any clear indication of the existence of massive stellar halos using S\'ersic index when galaxies are measured at low or moderate inclination angles.

    In summary, the physical origin of the large S\'ersic index of classical bulges remains poorly understood. A lower central concentration partly accounts for the lower S\'ersic index in TNG simulations. However, there is no physical basis for establishing such a correlation between the S\'ersic index of morphologically defined bulges and kinematically derived stellar halos, as a result, stellar halos. This result challenges the view regarding the correlation between classical bulges and super-massive black holes, in which mergers were expected to play a vital role \citep{1977egsp.conf..401T,2004ARA&A..42..603K,2009MNRAS.397..802H,2010ApJ...715..202H}

\section{Conclusions}\label{sm}

This study explores the structural differences between traditional morphological and kinematic decompositions of Sombrero-like galaxies. Using TNG50, a high-resolution cosmological simulation, we selected 270 Sombrero-like galaxies at redshift 0 with stellar masses \(M_\ast > 10^{10} M_\odot\) and stellar halo fractions \(0.3 < f_{\rm halo} < 0.6\). Sombrero-like galaxies are common in the local Universe, making up approximately 30\% and 60\% of less and more massive galaxies, respectively. 

Sombrero-like galaxies are challenging to be identified due to the presence of bars, spiral arms, and active star formation viewed in low or moderate inclinations. Morphology and color-based classifications thus are unreliable for distinguishing Sombrero-like galaxies from disk-dominated galaxies. 

Photometric decompositions were compared to kinematic decompositions. Morphological decomposition tends to overestimate the disk mass fraction significantly when galaxies are viewed at low or moderate inclinations. While edge-on decomposition in morphology can provide a rough estimate of the stellar halo mass fraction, it exhibits considerable scatter. Additionally, there is a systematic difference in the detailed radial distribution of these components compared to the results obtained from kinematic decomposition.

The S\'ersic index fails to distinguish classical bulges from pseudobulges in TNG simulations, with Sombrero-like galaxies resembling pseudobulges. No correlation was found between the S\'ersic index and large stellar halos, suggesting that morphology alone may not reflect halo presence. These results highlight the need for improved methods to accurately decompose and understand Sombrero-like galaxies. The differences in the merger histories between Sombrero-like galaxies and fiducial disk-dominated galaxies thus may introduce substantial uncertainties in our understanding of galaxy formation and evolution.

\begin{acknowledgements}
      The authors thank the constructive discussion with L. C. Ho and D. A. Gadotti. The authors acknowledge the support by the China Manned Space Program through its Space Application System, the Fundamental Research Funds for the Central Universities (No. 20720230015), the Natural Science Foundation of Xiamen, China (No. 3502Z202372006), and the Science Fund for Creative Research Groups of the National Science Foundation (NSFC) of China (No. 12221003). The TNG50 simulation used in this work, one of the flagship runs of the IllustrisTNG project, has been run on the HazelHen Cray XC40-system at the High Performance Computing Center Stuttgart as part of project GCS-ILLU of the Gauss centers for Supercomputing (GCS). This work is also strongly supported by the Computing Center in Xi'an. 
\end{acknowledgements}
\bibliography{wx}{}
\bibliographystyle{aa}
\end{document}